\documentclass{aa}  
\usepackage{caption}

\usepackage{graphicx}
\usepackage{txfonts}
\usepackage{natbib}
\usepackage{enumerate}
\usepackage{bm}

\usepackage{txfonts}
\begin{document} 
   \title{ISM conditions  in z$\sim$ 0.2  Lyman-Break Analogs.}

   \author{A. Contursi \inst{1}
          \and
          A.J.  Baker \inst{2} 
          \and
          S. Berta \inst{1,10}
          \and
          B. Magnelli \inst{3}
          \and
          D. Lutz \inst{1}
          \and
          J. Fischer \inst{4}
          \and
          A. Verma \inst{5}
          \and
          M. Nielbock \inst{6}
          \and
          J. Gr\'acia Carpio \inst{1}
          \and
          S. Veilleux \inst{7}
         \and
          E. Sturm \inst{1}
         \and
          R. Davies \inst{1}
         \and
          R. Genzel \inst{1}
           \and
          S. Hailey-Dunsheath \inst{8}
           \and
          R. Herrera-Camus \inst{1} 
           \and
          A. Janssen \inst{1}
          \and
          A. Poglitsch \inst{1}
          \and
           A. Sternberg \inst{9}
          \and 
          L. J. Tacconi  \inst{1}
          }

   \institute{Max-Planck-Institut f\"ur extraterrestrische Physik, Postfach 1312, 85741 Garching, Germany
              \email{contursi@mpe.mpg.de}
        \and
         Department of Physics and Astronomy, Rutgers, The State University of New Jersey, 136 Frelinghuysen Road, Piscataway, NJ
	 08854-8019, USA
         \and
         Argelander-Institut f{\"u}r Astronomie, Universit{\"a}t Bonn, Auf dem H{\"u}gel 71, 53121, Bonn, Germany
         \and
         Naval Research Laborratory, Remote Sensing Division, 4555 Overlook Ave SW, Washington, DC 20375, USA
         \and
         Oxford University, Dept. of Astrophysics, Oxford OX1 3RH, UK
         \and
         Max-Planck Institut f{\"u}r Astronomie, K{\"o}nigstuhl 17, 69117  Heidelberg, Germany; Universit{\"a}t Heidelberg, 
         Zentrum  f{\"u}r Astronomie, Inst. f{\"u}r  Theor. Astrophysik, Albert-Ueberle-Str. 2, 69120  Heidelberg, Germany 
         \and 
         Department of Astronomy and Joint Space Science Institute, University of Maryland, College Park, MD 20742, USA 
         \and      
         California Institute of Technology, 301-17, 1200 E. California Blvd., Pasadena, CA 91125, USA        
          \and
         Tel Aviv University, Sackler School of Physics $\&$ Astronomy, Ramat Aviv 69978, Israel
          \and
         Department of Physics, Faculty of Science, University of Zagreb, 
         Bijeni\v{c}ka cesta 32, 10000  Zagreb, Croatia \thanks{visiting scientist}\\
         }
 
    \abstract
    {We present an analysis of   far--infrared (FIR) [CII] and [OI] fine structure line   and  continuum observations 
    obtained with $Herschel$/PACS, and $^{12}$CO(1-0)  observations obtained with the IRAM Plateau de Bure Interferometer, 
    of Lyman Break Analogs (LBAs) at $z\sim 0.2$.  
    The principal aim of this work is to determine the typical ISM  properties  of $z\sim 1-2$ Main Sequence (MS)
    galaxies, with stellar masses between $10^{9.5}$ and $10^{11}$ $M_{\odot}$,
    which are currently not easily detectable in all these lines even with ALMA
    and NOEMA.  \\
    We perform PDR modeling and apply different IR diagnostics to derive the main physical 
    parameters of the FIR emitting gas and   dust and we compare the derived ISM properties
     to those of galaxies on and above the MS at different redshifts.
    We find that the ISM properties of LBAs are quite extreme (low gas temperature, high density and thermal pressure)
    with respect to those  found in local normal spirals and more  active local galaxies.
    LBAs have no [CII] deficit despite having the high specific star formation rates (sSFRs)
    typical of starbursts. 
    Although LBAs lie above the local MS,  we show that their ISM properties are more similar to those
    of high-redshift MS galaxies than of local galaxies above the main
    sequence.\\ 
    This data set represents an important reference for planning future ALMA [CII] observations of 
    relatively low-mass MS galaxies at the epoch of the peak of the cosmic star formation. }

    \maketitle

\section{Introduction}
 In the last decade, increasing observational and theoretical studies at low and high redshift 
have shown that the conditions of the Interstellar Medium (ISM) in galaxies have a primary role in their evolution.
In the stellar mass ($M_*$) $vs.$  star formation rate (SFR) plane, star-forming galaxies (SFGs) 
 at all redshifts  sampled so far  \citep[up to $z \sim$ 6,][]{Salmon} follow a tight correlation called the Main Sequence (MS) such that  the
SFR  increases with  redshift  for a given stellar mass.
Nevertheless, at a given stellar mass, the efficiency of forming stars    
({\it i.e.}, the SFR per unit of  molecular gas mass) is broadly constant for MS 
galaxies at all redshifts, and the increase of the SFR with $z$
is primarily driven by an increasingly high  gas fraction 
\citep{Daddi10b,Saintonge11,Tacconi10,Tacconi13,Genzelscaling}.\\  
Observations have also shown that the gas distribution and dynamical state of MS galaxies evolves with redshift.
Kinematic and dynamical studies of the ionized and molecular gas in MS SFGs
at the epoch where the cosmic star formation peaks, i.e., $z$=2-3, have shown that by this 
time the more massive MS galaxies have already formed regularly rotating disks similar to those of local star-forming galaxies. 
However, the gas is clumpier and  has a higher velocity dispersion  than the gas in local MS galaxies
\citep{Wisnioski}. 
Thus, the ISM dynamical state, the gas content,  and the rate at which a galaxy is turning gas into stars 
are amongst the fundamental parameters that regulate the evolution of a galaxy.\\
\noindent  
Most  past studies of the ISM in high-redshift galaxies have analyzed
 ionized and molecular gas. The cold neutral atomic medium is another important  phase of the ISM in galaxies,
 and its study can provide complementary  information on the physical conditions in the ISM and hence on   galaxy evolution.
One way to study this phase of the ISM is with  observations of the  [CII] ($^2$P$_{1/2}$$\rightarrow$$^2$P$_{3/2}$) line at 157.7 $\mu$m and the 
 [OI] ($^3$P$_2$$\rightarrow$$^3$P$_1$) line at 63.2 $\mu$m, the  
  most luminous far-infrared (FIR) fine structure lines: in  local SFGs these lines
  represent  0.1-1$\%$ of the FIR luminosity  \citep{Malhotra,Javier11}.  
These lines are the major coolants of the ISM, and therefore they are very important
for the energetic balance of the ISM. \\
The recent availability  of  new sensitive submillimeter interferometers has opened the possibility of observing
[CII] in  high-redshift  galaxies \citep{Carilli} where  observation of the HI hyperfine line at 21 {\rm cm} 
is not yet possible. However, these observations are still challenging at $z = 2-3$, and so far restricted
to the most luminous and massive systems (QSOs, submillimeter galaxies, massive MS galaxies). 
What is ultimately  needed is the study of the cold atomic gas in   high-redshift  
MS SFG samples matching those already  studied in  
ionized and molecular gas.\\
An ideal sample is represented by the Lyman break galaxies  \citep[LBGs,][]{LBGs},  
 selected using color criteria sensitive to the presence of a Lyman continuum break in an
otherwise very blue rest far-UV continuum in the redshift range $z\sim 2-3$. LBGs have been extensively studied
at many  wavelengths and represent an obvious population to be observed in the [CII] line. 
However, observations of such relatively low-infrared-luminosity systems at high redshift  are still very 
difficult even with the state--of--the--art receivers of modern submillimeter interferometers.\\
In order to study the ISM conditions in such systems in more detail,  we  have undertaken 
a   program to observe  a small sample of local  ($z\sim$ 0.2) analogs of LBGs,
 ({\it {Lyman--break analogs}} or LBAs; 
 \citep{Heckman05,Heckman11,Hoopes07,Overzier08,Overzier09,Overzier10,Overzier11,Basu-Zych07,Basu-Zych09,LBAHST} in [CII] and [OI] line emission 
 as well as in  FIR continuum  emission. Our study uses   the PACS spectrometer and photometer \citep{Albrecht} 
 on--board the $Herschel$ satellite  \citep{Pilbratt}  as part  of a guaranteed--time program 
(P.I. A. Contursi). We also present  $^{12}$CO(1-0) IRAM Plateau de Bure (PdBI) observations of some of our targets in order to
extend our analysis to their molecular gas.   
With this work,  we aim also to provide a unique
 reference for designing  ALMA [CII] and continuum observations of  SFGs at the epoch where cosmic star
 formation peaks.\\
The paper is organized as follows. In Section 2, we introduce the sample. In Section 3, we describe the data reduction and the
extraction of the line and continuum fluxes. The derivation of the main parameters  necessary for the analysis is given in
Section 4. In Section 5, we present the main results obtained by modeling the FIR fine structure
lines. In Section 6, we discuss these results  in the
framework of  galaxy evolution. A summary of the results and conclusions is given in Section 7.

\begin{figure}
\centering
  
\includegraphics[angle=0,width=10.0cm,height=6.0cm]{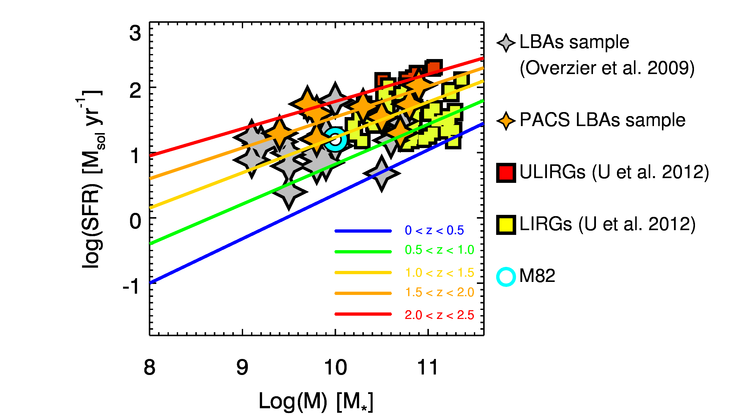}
      \caption{The location of LBAs on the SFR-M$_*$ plane. Diamonds correspond to the   sample of 
      \citet{Overzier09}. Orange diamonds represent the subsample presented in this work. SFRs and stellar masses are from 
      \citet{Overzier09}. SFRs are calculated from  H$\alpha$ + 24 $\mu$m flux assuming a Kroupa IMF. The sample of ULIRGs (red) 
      and LIRGs (yellow) is taken from  \citet{U}. 
      The overplotted lines are the relationships followed by star-forming galaxies at different redshifts
       presented in \citet{Whitaker12}, who use a similar derivation for the SFR. The 
       position of  the typical local starburst galaxy
        M82 is also shown for reference.}
         \label{LBA_and_the_others}
   \end{figure}

  \section{The sample}
\subsection{Lyman Break Analogs}
Lyman Break Analogs (LBAs) are   ultra-compact UV luminous galaxies at redshift
$z$$\sim$ 0.1-0.3 selected to have 
$L_{UV} > 2 \times 10^{10}$ $\rm{L_{\odot}}$  and very high UV
surface brightness $\rm{I_{1530~\AA}}  > 10^9$
$\rm{L_{\odot}~ kpc^{-2}}$  \citep{Heckman05}.  
With UV half-light radii between 0.5 and a few kpc  \citep{Overzier10}, 
they are more compact and have higher SFR per unit area than  large UV--bright local spirals \citep{Hoopes07}. 
 They are more luminous than Blue Compact Dwarfs
in the UV and much more rare.  Their SFRs span   from $\sim 3$ to 100  $M_\odot~yr^{-1}$, and their specific SFRs (sSFRs) are
higher than those of  large UV--bright galaxies. 
  \citet{Overzier09} propose   that LBAs   could represent  a very short, relatively unobscured evolutionary phase of a
starburst.
Several of these galaxies  show disturbed morphologies and tails,
indicating signs of interactions, with generally disk like structures featuring compact luminous 
clumps. Some LBAs are dominated by  very UV--bright, compact ($\sim$ 10$^2$ pc), and massive ($\sim$
10$^9$ M$_\odot$) clumps. These galaxies are called {\it "Dominant Central Objects"} (DCOs)  
and exhibit  higher sSFR than the rest of the sample  
\citep{Overzier09}. Some  DCOs also show  clear signs of high-velocity ionized gas outflows ( $\sim 400$  {\rm km s$^{-1}$} < v < 1500 
 {\rm km s$^{-1}$}) 
and a high percentage (up to $\sim$
40$\%$) of escaping Lyman continuum photons thanks to a clumpy neutral medium  \citep{Heckman11}. \\  
\noindent
LBAs and LBGs share many properties, including their ranges of UV luminosity and surface brightness, 
morphology,  stellar mass, SFR and metallicity. 
 LBAs, like LBGs, suffer   less attenuation than local galaxies with the same bolometric
 luminosities  \citep{Overzier11}: in the IR/UV $vs.$ the stellar mass plane, LBAs 
  lie
 well below the relation followed by local SDSS galaxies. 
LBAs also follow three fundamental scaling relations that further show their similarity to LBGs. 
First, in the stellar mass --  metallicity relation shown in Figure 9 of \citet{Hoopes07} \citep[see also][]{Lian},  
LBAs  follow the  mass-metallicity relation  of $z$ $\sim$ 2    UV star-forming galaxies    \citep{Erbb}.
 There is also some overlap between LBAs and local (Ultra) Luminous Infrared Galaxies ((U)LIRGs),  
 but LBAs have on average lower metallicities (sub-solar) than (U)LIRGs  
  \citep[1-2 $\times$ $Z_{\odot}$][]{Inami}  and lower extinction. 
Second, \citet{Overzier09} show  that LBAs have [N\ II]/H$\alpha$
  ratios  higher than   those of nearby warm IR galaxies with the same [O\ III]/H$\beta$ ratio 
  \citep[BPT]{BPT,BPT2}, but similar  to the ratios of  a variety of $1 < z <   2.5$ 
  star-forming galaxy samples  \citep{BPThighz}.   This suggests   
  similar  conditions in the ionized gas for   LBAs and high--$z$ galaxies. 
  The LBAs that occupy the composite HII-AGN region 
  of the BPT diagram are also DCOs.
  There is no clear evidence that these galaxies   host    hidden AGN,
  except   for  SDSS092159, which most likely hosts a black hole with a mass $\leq 10^7$ $M_{\odot}$  \citep{Alexandroff}.
  This does  not exclude the presence of  low-luminosity AGN in the other composite 
  sources, but   if   AGNs are present,   they do  not dominate the emission.  
Third, Figure  \ref{LBA_and_the_others} shows the original sample  of LBAs from  \citet{Overzier09} 
in the SFR-$M_*$ plane.
The subsample of galaxies presented in this work is shown in orange diamonds (see next section for the sample selection
criteria). For comparison, we plot a sample of ULIRGs and LIRGs  from  \citet{U}, the position 
of the local starburst M82, and the  main sequence at different redshifts  as published by  \citet{Whitaker12}.
 LBAs lie above the local SFR-M$_*$ relation followed by normal star-forming galaxies. 
 The LBAs we have selected to be observed with PACS  are LIRGs, and in fact they overlap in part with the 
 LIRGs sample of \citet{U}, but they are on average less massive and have lower metallicities and IR/UV ratios. \\

\subsection{Target selection}
We selected  nine LBAs  from the Overzier et al. (2009) sample based on two criteria: 
(1) the  redshift must be 
such that the [CII] line falls at a wavelength lower than  196
$\mu$m, beyond which the leakage from the second order makes the flux calibration very
uncertain, and (2)  the IR continuum emission must be
bright enough to reach S/N $\gtrsim 10$ in a reasonable amount of time with PACS  
(not more than 3 hours  per line per source). \\
Table \ref{Table1} lists the selected targets with their 
main characteristics  from   \citet{Overzier09} and   \citet{Basu-Zych07}.
 We observed all nine targets in 
[CII] and five of   the  nine in the 63 $\mu$m  [OI] line. 
We   observed all nine targets with the  
PACS photometer in the green (100 $\mu$m) and red (160 $\mu$m) broadband filters.
Since LBAs have been extensively studied at many wavelengths in previous works, 
many of their properties are already known for all targets. We  summarize the most important of these in Table \ref{Table2}.\\
Figure  \ref{LBA_and_the_others}  shows that our sample (orange diamonds), despite being small 
and not complete,  spans most of the parameter space covered by 
the  original sample of  \citet{Overzier09}, (gray diamonds), although is  biased towards
slightly higher SFR and stellar mass.\\
By complementing the FIR data  with IRAM PdBI $^{12}$CO(1-0) observations  of four targets, presented here for the
first time, and with   CARMA $^{12}$CO(1-0)  data   available for two additional galaxies from  
\citet{LBACO},  we are able to probe the neutral atomic and molecular gas and dust content of LBAs.\\

\begin{table*}
\caption{The sample observed with the PACS photometer and spectrometer. }               
\begin{tabular}{clccccccccl}
\noalign{\smallskip}
\hline
\noalign{\smallskip}
 $\#$ &   Name  &  R.A.     &       Dec.    &    z    &   logM$_*$     &  Z$^+$     &        SFR                &   logL$_{24}$     &    logL$_{70}$    &  \\
  &         &  J2000        &      J2000    &         &   (M$_\odot$) &            &    (M$_{\odot}$/ yr)       &   (L$_{\odot}$)   &   (L$_{\odot}$)   &  \\
  &         &               &               &         &               &            &    H$\alpha$+24$\mu$m      &                   &                   &  \\
\noalign{\smallskip}
\hline
\noalign{\smallskip}
 1 & 005527           &   00:55:27.46  & -00:21:48.71 & 0.16744**  &         9.7   &    8.28    &          55.4             &      10.8        &       11.12       &  \\      
 2 & 015028$^\dagger$ &   01:50:28.41  &  13:08:58.40 & 0.14668**  &   10.3   &    8.39    &               50.7             &      10.8        &       11.47       &  \\
 3 & 021348$^\dagger$ &   02:13:48.54  &  12:59:51.46 & 0.21902**  &   10.5   &    8.74    &               35.1             &      10.7        &        ---        &  \\
 4 & 080844           &   08:08:44.27  &  39:48:52.36 & 0.09123**  &         9.8   &    8.74    &          16.1             &      10.3        &       10.53       & \\
 5 & 082001           &   08:20:01.72  &  50:50:39.16 & 0.217      &    9.8   &    8.15    &               40.0             &      10.6        &       11.00       & \\
 6 & 092159           &   09:21:59.39  &  45:09:12.38 & 0.23499**  &   10.8   &    8.67    &               55.1             &      10.8        &       11.60       & \\
 7 & 093813           &   09:38:13.50  &  54:28:25.09 & 0.10208**  &         9.4   &    8.19    &          19.8             &      10.2        &       10.54       &  \\
 8 & 143417$^\dagger$ &   14:34:17.16  &  02:07:42.58 & 0.180      &        10.7   &    8.65    &          20.0             &      10.1        &       10.86       & \\
 9 & 210358$^\dagger$ &   21:03:58.75  & -07:28:02.45 & 0.13698**  &        10.9   &    8.70    &         108.3             &      11.1        &       11.35       & \\
\noalign{\smallskip}

   \hline
\multicolumn{9}{l}{Data are from  \citet{Overzier09} 
except  for updated redshifts when available ($**$) 
 from   \citet{Heckman11}}\\

\multicolumn{9}{l}{$^+$ 12+log(O/H) estimated using the O3N2 estimator from  \citet{Pettini}.} \\     
\tiny{$^\dagger$ CO available}  &           &       &             &                  &                           &                 &                         &              &     \\     
 
\hline
\noalign{\smallskip}
\label{Table1}
\end{tabular} 
\end{table*}

\begin{table*}
\caption{Classification and morphological properties of the LBA  sample  presented in this work.
We also report the percentage of the Lyman continuum leakage that indicates the presence of a  clumpy neutral phase, that we  study in this work. }               
\begin{tabular}{clcccccccccccccccccccccc}
\noalign{\smallskip}
\hline
\noalign{\smallskip}
 Name  & DCO$^{1}$  &   Interacting      &  Ionized gas  $v_{max}^{Outflow}$ $^3$ &  High $\%$ escape   &    BPT  $^1$                   &         AGN$^1$          &      \\
       &            &   /merger$^2$      &        ${\rm km ~ s^{-1}}$             &     fraction  $^3$                               &       classification              &       signatures       &      \\
  
\noalign{\smallskip}
\hline
\noalign{\smallskip}
005527     &   no      &  no            &              -640                      &                    no                   &       HII          &           no               &     \\      
015028     &   no      &  tidal ?       &              -480                      &                    no                   &       HII          &           no               &     \\
021348     &  yes      &  --            &              -1500                     &                   yes                   &       Comp.        &           no               &     \\
080844     &  yes      &  companion     &              -1500                     &                   yes                   &       Comp.        &           no               &     \\
082001     &   no      &   no           &                --                      &                    no                   &       Comp.        &           no               &     \\
092159     &  yes      &   --           &              -1500                     &                    no                   &       HII          &          yes               &     \\
093813     &   no      &   merger       &               -610                     &                   yes                   &       HII          &           no               &     \\
143417     &   no      & interaction    &                --                      &                    no                   &       Comp.        &           no               &     \\
210358     &  yes      &   no           &              -1500                     &                    no                   &       Comp.        &           no               &     \\
\noalign{\smallskip}

   \hline

\tiny{$^1$  \citet{Overzier09}}     &      &            &                          &                  &                                         &                   &                            &      \\
\tiny{$^2$  \citet{LBAHST}}         &      &            &                          &                  &                                         &                   &                            &      \\
 
\tiny{$^3$  \citet{Heckman11}}      &      &            &                           &                  &                                         &                   &                            &      \\
\hline
\noalign{\smallskip}
\label{Table2}
\end{tabular} 
\end{table*}

   \begin{figure*}
   \centering
\includegraphics[angle=0,width=20.5cm,height=18.5cm,trim=0.7cm 5cm 0cm
2cm,clip]{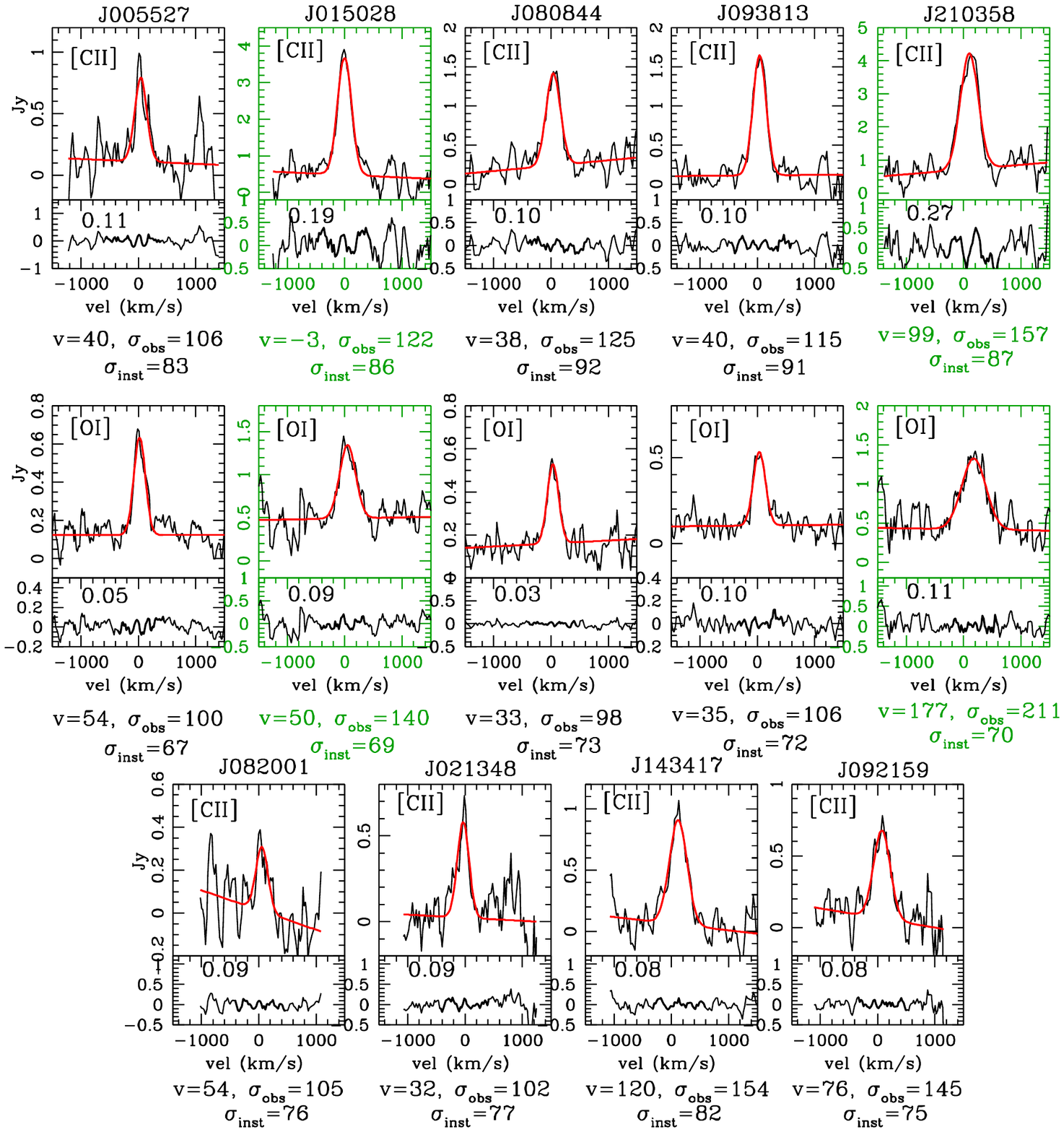}
        \caption{ PACS  spectra of all targets in {\rm km s$^{-1}$}. Zero velocity corresponds to the redshift  listed in Table \ref{Table1}.
     Each black or green  frame contains the single   Gaussian fit for [CII] (upper) and [OI] (lower when available)
  and residuals for each target. We report the  the fitted peak velocities, velocity dispersion  $\sigma_{obs}$ and  
  instrumental $\sigma_{inst}$  at the observed wavelength in {\rm km s$^{-1}$}. Black frames refer to spectra extracted 
  from the central spaxel with the Point Source Correction; green frames refer to extraction from the sum of the 
  central 9 spaxels with the Point Source Correction, to correct for line distortion introduced in badly pointed data (see text for details). }
         \label{Spectrakms}
   \end{figure*}

\section{Observations and data reduction}
\subsection{PACS data}
We list the details of all PACS observations in Table \ref{Table3}. 
All spectrometer observations were executed in chop--nod line (small range) scan mode for [CII] ([OI]). 
The [OI] integration times were typically 4 times longer than those of [CII], often approaching one hour. 
The spectrometer data have been reduced in HIPE{\footnote{HIPE is a joint development by the Herschel Science Ground
Segment Consortium, consisting of ESA, the NASA Herschel Science
Center, and the HIFI, PACS and SPIRE consortia.}} using the telescope background normalization method. This method
does not use the  calibration block executed at the beginning of the observations to flux--calibrate the spectrum
but instead uses the telescope background spectrum itself. The advantage is that in this case, each   chop on-off calculation   
is normalized to the sum  of chop  on  plus chop  off, resulting in a   signal   normalized to the telescope
background emission for every chopping cycle, thereby correcting the responsivity drifts much more efficiently and at  much higher
temporal sampling than is done with the calibration block reduction.
This  method is particularly recommended for long observations  (such as ours), and 
it also  usually works better in recovering  baselines  when  targets are very faint or,
for   point sources, where the continuum flux is expected to be zero. 
We have verified that our targets are all   point sources at all wavelengths with respect to  the PACS  beams (FWHM from 9 to 13") 
in agreement with  their UV and optical extension of
$\lesssim 5\arcsec$  \citep{LBAHST}. \\

\noindent
To extract a point source  flux, there are two methods. One consists of extracting
the spectrum from the central spaxel only and   applying a point source  correction loss
 to recover the total  flux. This
 offers the advantage that the spectrum of the central spaxel has the highest S/N. However, if the observation
 is significantly  mispointed ($\gtrsim 3"$),  one can sum the signal of the central 3$\times$ 3 spaxels and
 apply a correction from this to the total flux, thus reducing the effect of any possible
 mispointing. The disadvantage with this method is that  if the signal is faint,  as   is the case for most of our
 targets, summing   the central nine spaxels  increases the noise.\\
 Given the latter consideration, we have chosen to extract the final source spectra using the  central spaxel for all targets 
 except when the mispointing is evident from the 5x5 spaxel  flux distribution.  
This is the case  for  two galaxies, SDSSJ210358 and SDSSJ015028, in which the 
 [CII] and [OI] lines, which should arise in the same gas, appear spectrally misaligned. 
 In fact, when a target is mispointed in a direction perpendicular to the slices,   spectra are  artificially 
 skewed, and  thus  their line profiles, 
 and hence  velocity peaks and widths, are altered (see Figure 14.18 of the PACS spectroscopy Calibration Document\footnote{
http://herschel.esac.esa.int/twiki/pub/Public/PacsCalibrationWeb/\\
PacsSpectroscopyPerformanceAndCalibration\_v2\_4.pdf}). 
For these two galaxies, we extract  the sum of the 3x3 central spaxels to recover the line profiles more correctly.\\  
 The  nominal flux uncertainties ($\sim$ 10$\%$) reported in the PACS spectrometer calibration
 document  are  strictly valid only for
 continuum flux $\gtrsim 1 $  Jy. In the much fainter regime of emission of our targets, uncertainties are
 likely to be higher. Taking all these facts together, we estimate our flux uncertainties for the lines  
 to be  $\sim$25$\%$.\\
 
All photometer observations were executed in small scan mode with two scan directions in the green and red filter. 
We reduced them with the standard
pipeline and extract fluxes by using aperture photometry and applying the point source corrections.
We have also derived the continuum flux densities from the spectrometer data on either side of the  targeted
lines and in the corresponding parallel channels. The errors on the continuum fluxes measured from the spectra are in 
general higher than those  at the line centers because the sampling is lower there. 
Table \ref{Table4} lists the continuum flux densities derived from  the photometer and the spectrometer
 data.

\begin{table}[h]
\caption{PACS photometer and spectrometer observational details. }               
\begin{tabular}{clcccccccccccccccc}
\noalign{\smallskip}
\hline
\noalign{\smallskip}
  Name  &  Line/band     &  OBSID    &    Time  on source  &\\ 
      & 	      & 	       &       (sec)	     & \\ 
      & 	      & 	       &		     & \\ 
\noalign{\smallskip}
\hline
\noalign{\smallskip}
  005527    &  [CII]	 &  1342213130  	    & 860     &\\ 
	  &  [OI]      &  1342213315		    & 3920    &\\ 
	  &  Green+red     &  1342222493/4   & 180   &\\ 
\noalign{\smallskip}
 015028    &  [CII]	&  1342213912		   & 860     & \\
	 &   [OI]     &  1342213750		 &  3920     & \\
	 &  Green+red	  & 1342223572/3    & 180   & \\
\noalign{\smallskip}
  021348    &  [CII]	 &  1342213911  	    & 3440    & \\
	  &  Green+red     & 1342223654/5    & 360   & \\
\noalign{\smallskip}
 080844    &  [CII]	& 1342220752		   & 860     &\\ 
	 &  [OI]      & 1342220141		 & 2800      &\\ 
	 &  Green+red	  & 1342220125/6    & 180   &\\ 
\noalign{\smallskip}
 082001    &  [CII]	& 1342220751		   &  3440  &\\ 
 	&  Green+red	 & 1342220121/2    & 360   &\\ 
\noalign{\smallskip}
  092159    & [CII]	 & 1342221362		    &	3440  &\\ 
	  &  Green +red    & 1342220657/8    & 180   &\\ 
\noalign{\smallskip}
   093813    & [CII]	  & 1342221632  	     & 860    &\\ 
	  & [OI ]      & 1342221631		  & 2800      & \\
	  &  Green+red     & 1342220661/2    & 180   & \\
\noalign{\smallskip}
  143417    & [CII]	 & 1342213757		    & 3440    & \\
	  &  Green+red     & 1342223838/9    & 360   &\\ 
\noalign{\smallskip}
   210358    & [CII]	  & 1342208937  	     & 860    &\\ 
	  & [OI]       & 1342221358		  & 3840      &\\ 
	  &  Green+red     & 1342218538/9     & 180  &\\ 
\noalign{\smallskip}
 \hline
\noalign{\smallskip}
\label{Table3}
\end{tabular} 
\end{table}

\begin{table}

\caption{Continuum flux densities and uncertainties calculated from the maps observed with the PACS photometer (in bold and 
with superscript $p$), and
 from the continuum  values on both sides of  the lines and from the parallel channels of the spectrometer observations ($s$). 
The uncertainties are the  measured  {\it r.m.s.}.              
}              
\tiny
\begin{tabular}{cccl}
\noalign{\smallskip}
\hline
\noalign{\smallskip}
    Name  &$  \lambda$  &  Flux   density      &   \\
          &    $\mu$m   &          mJy         &   \\
          &             &                      &   \\
&              &                      &   \\

\noalign{\smallskip}                                                                                            
\hline                                                                                                          
 \noalign{\smallskip}                                                                                           
005527  &                       &                         &\\                                                   
        &  73.76$^{s}$          &  152.5         $\pm$   78.7   &\\                                                             
        &  92.11$^{s}$          &  178.0         $\pm$  145.0   &\\                                                             
        & {\bf 100.00$^{p}$ }   &  {\bf 109.9    $\pm$   10.9 }  &\\       
        & 147.52$^{s}$          &   77.3         $\pm$   25.9   &\\       
        & {\bf 160.00$^{p}$ }   &  {\bf  76.7    $\pm$   18.2 }  &\\                                                            
        & 184.17$^{s}$          &  103.7         $\pm$  193.1   &\\                                                     
\hline                                                                                                          
 \noalign{\smallskip}                                                                                           
015028  &                       &                         &\\                                                   
        &  72.50$^{s}$          &  494.6         $\pm$   78.8   &\\                                                     
        &  90.49$^{s}$          &  575.9         $\pm$  146.2   &\\                                                             
        & {\bf 100.00$^{p}$}    &  {\bf 487.4    $\pm$   20.3}   &\\                                                    
        & 145.03$^{s}$          &  339.6         $\pm$   23.0   &\\                                                             
        & {\bf 160.00$^{p}$}    &  {\bf  334.1   $\pm$   14.2}   &\\                                                            
        & 180.87$^{s}$          &  404.7         $\pm$  192.837 &\\                                                             
\hline                                                                                                          
\noalign{\smallskip}                                                                                            
021348  &  96.18$^{s}$          &  145.4         $\pm$   83.2   &\\                                                             
        & {\bf 100.00$^{p}$ }   &  {\bf 302.36   $\pm$    5.5}   &\\                                                    
        & {\bf 160.00$^{p}$ }   &  {\bf 211.2    $\pm$   16.3 }  &\\                                                            
        & 192.28$^{s}$          &   91.2         $\pm$   98.4   &\\                                                             
 \hline                                                                                                         
\noalign{\smallskip}                                                                                            
080844  &  68.95$^{s}$          &  145.7         $\pm$   70.6   &\\                                                             
        &  86.09$^{s}$          &  270.9         $\pm$  161.7   &\\                                                             
        & {\bf 100.00$^{p}$ }   &  {\bf 200.7    $\pm$  135.1}   &\\                                                    
        & 137.95$^{s}$          &  200.5         $\pm$   33.9   &\\                                                     
        & {\bf 160.00$^{p}$ }   &  {\bf 146.3    $\pm$  110.2}s  &\\                                                            
        & 172.10$^{s}$          &  262.2         $\pm$  143.7   &\\                                                             
\noalign{\smallskip}                                                                                            
\hline                                                                                                          
082001  &  96.04$^{s}$          &   64.6         $\pm$  119.1   &\\                                                             
        & {\bf 100.00$^{p}$}    &  {\bf  68.7    $\pm$   10.0}   &\\                                                    
        & {\bf 160.00$^{p}$}    &  {\bf  67.54   $\pm$    5.0}   &\\                                                            
        & 191.99$^{s}$          &    4.7         $\pm$  116.603 &\\                                                             
\noalign{\smallskip}                                                                                            
 \hline                                                                                                         
092159  &  97.41$^{s}$          &   61.2        $\pm$  125.8   &\\             
        & {\bf  100.00$^{p}$ }  &  {\bf 264.59  $\pm$    7.7}   &\\  
        & {\bf 160.00$^{p}$  }  &  {\bf 214.09  $\pm$   11.5}   &\\     
        & 194.75$^{s}$          &   95.4        $\pm$  101.6   &\\                                                              
\noalign{\smallskip}                                                                                            
 \hline                                                                                                         
093813  &  69.66$^{s}$         &  102.2         $\pm$   65.3   &\\                                                      
        &  86.95$^{s}$         &  195.9         $\pm$  159.4   &\\                                                              
        & {\bf 100.00$^{p}$}   &  {\bf  130.3   $\pm$    6.8}   &\\                                                     
        & 139.35$^{s}$         &   81.3         $\pm$   14.8   &\\                                                              
        & {\bf 160.00$^{p}$}   &  {\bf  77.0    $\pm$    7.1}   &\\                                                             
        & 173.82$^{s}$         &  115.5         $\pm$  137.2   &\\                                                              
\noalign{\smallskip}                                                                                            
 \hline                                                                                                         
143417  &  93.14$^{s}$         &   76.3        $\pm$   82.51  &\\                                                               
        & {\bf 100.00$^{p}$}   &  {\bf 103.9   $\pm$    8.3}   &\\   
        & {\bf 160.00$^{p}$}   &  {\bf  89.0   $\pm$   15.0}   &\\                                                      
        & 186.26$^{s}$         &   47.6        $\pm$  110.809 &\\                                                               
\noalign{\smallskip}                                                                                            
 \hline                                                                                                         
210358  &  71.87$^{s}$         &   419.9       $\pm$  102.5   &\\                                                       
        &  89.697$^{s}$        &   621.7       $\pm$  152.3   &\\                                                               
        & {\bf 100.00$^{p}$}   &   {\bf 603.8  $\pm$    8.2}  &\\                                                       
        & 143.80$^{s}$         &   567.1       $\pm$   18.3   &\\                                                       
        & {\bf 160.00$^{p}$}   &   {\bf 508.4  $\pm$   26.2}   &\\                                                              
        & 179.32$^{s}$         &   551.2       $\pm$  147.6   &\\                                                       
                                                                                                                
\noalign{\smallskip}                                                                                            
\label{Table4}                                                                                                  
\end{tabular}                                                                                                   
\end{table}

\subsubsection{[CII] and [OI] fluxes and line profiles}
Figure \ref{Spectrakms} shows the observed spectra as a function of velocity in ${\rm km ~ s^{-1}}$,
relative to the optical systemic velocity listed in column 5 of Table 1.
Both [CII] and [OI] lines were detected in all  galaxies  except  in  SDSSJ082001, which shows a marginal detection in   [CII] 
(it was not observed in [OI])  that we treat as upper limit.\\
The spectra were  extracted as explained in section 3.1. 
For each target we show  fits to a  single Gaussian plus   linear  baseline model   and   the corresponding residuals at the bottom 
of each spectrum. In each residual panel, we show the standard deviation calculated in the spectral region  indicated 
with a thicker line. We also report the fitted velocity peak, the observed line dispersion  and the instrumental dispersion
at the observed wavelength in ${\rm km ~ s^{-1}}$. 
Taking into account all uncertainties due to the nominal wavelength calibration and 
to the line profile changes introduced by mispointing, we estimate that the  overall uncertainty in  
velocity is $\sim$ 50  ${\rm km ~ s^{-1}}$ in both lines. 

\begin{figure}
\centering
\includegraphics[angle=0,width=9.5cm,height=7.5cm,trim=2.5cm 14.5cm 6cm
4.5cm,clip]{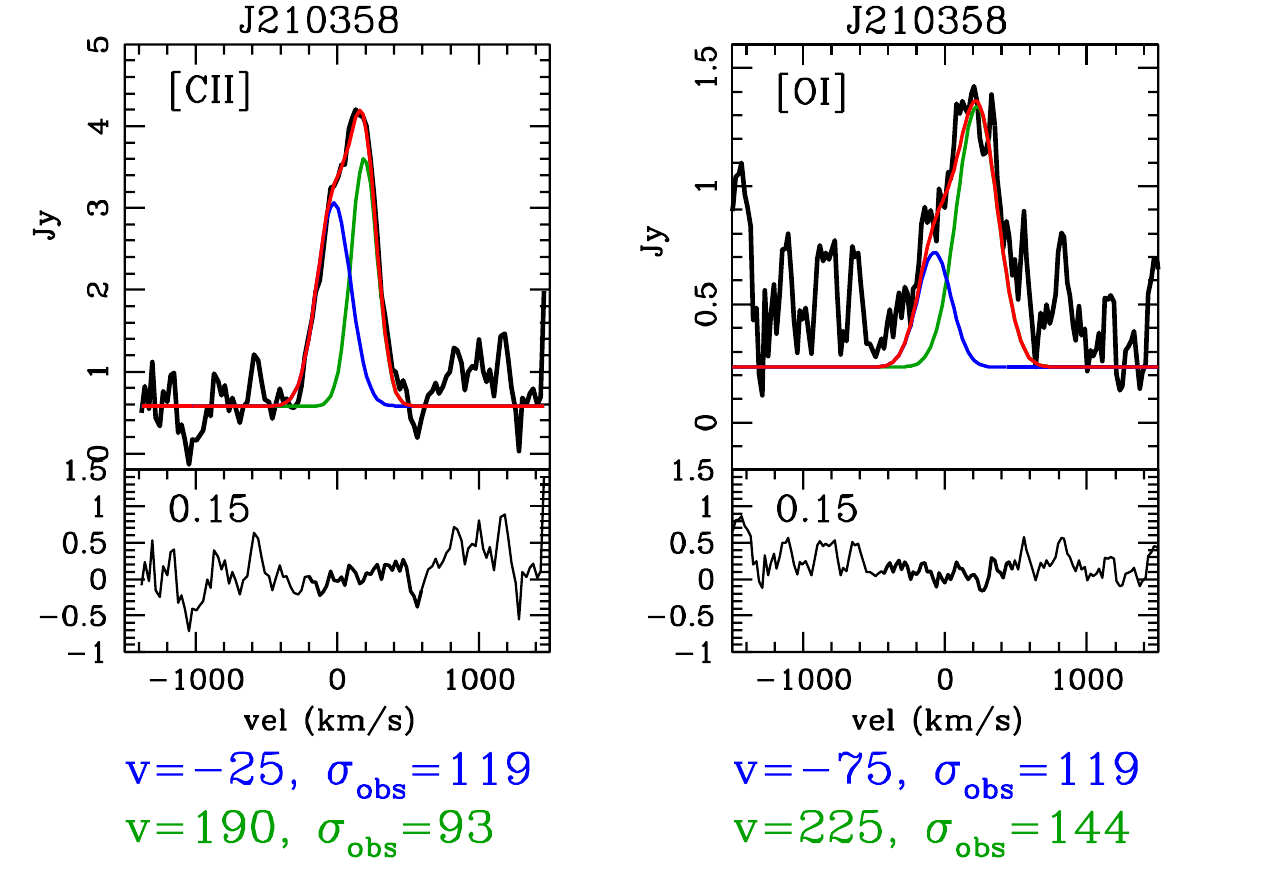}
       \caption{ [CII] and [OI] spectra of SDSSJ210358, with two   Gaussian component  plus polynomial continuum fit, which yields   
       residuals smaller than a fit with one Gaussian component.}
         \label{SpecialCaseSDSSJ210358}
\end{figure}

\noindent
The fitted parameters of the two lines   marginally agree for galaxy
SDSSJ015028 but differ by more than the uncertainties for galaxy SDSSJ210358.
The spectra of the latter galaxy clearly show two components, so we fit both [CII] and [OI] with two Gaussian components 
plus a linear continuum, and we obtain better agreement between the peak
velocities  for both components for the two lines. This is shown in Figure \ref{SpecialCaseSDSSJ210358}, where it is also clear that the
dispersion of the residuals is much lower than that obtained in the case of one fit component.

\noindent
For completeness, we have also performed double Gaussian fits for all targets, 
to explore the presence of multiple or broad components, but we do not obtain improvement in the fits.
This result, however, does not rule out the existence of broad wings indicating  outflowing neutral 
atomic gas,  given the low S/N.
Table \ref{Table5} lists the line flux, the S/N, and  the instrumental and observed FWHMs derived from the single-Gaussian fits.\\
 We clearly resolve the lines in SDSSJ143417, SDSS092159,
 SDSSJ210358 and   SDSSJ015028. The other targets show line widths that are consistent with the instrumental spectral 
 resolution  or marginally resolved.

\subsection{$^{12}$CO(1-0) data}
We obtained IRAM  PdBI   
 \citep{IRAM} observations of four LBAs in the array's compact D
configuration between 2005 August and 2007 June.  The four
targets were selected from the  \citet{Heckman05} sample on
the basis of significant detections in 1.2\,mm photometry with
the Max-Planck Millimeter Bolometer (MAMBO) array on the IRAM
30\,m telescope and in at least one {\it IRAS} band, on the
assumption that dust emission would augur well for a gas
detection experiment.  Observations of SDSSJ015028 (five tracks
with 4-5 antennas in operation) and SDSSJ143417 (three tracks with
5 antennas) were obtained in 2005--06 through program O060;
observations of SDSSJ021348 (two track with 4--5 antennas) and
SDSSJ210358 (three tracks with 4--6 antennas) were obtained through
program Q073.  All observations were calibrated with the
CLIC package in the GILDAS environment  
\citep{IRAM2}, using nearby quasars for gain calibration, bright but
more distant quasars for passband calibration, and the radio
star MWC349 (or a bright quasar bootstrapped to MWC349) for
flux calibration.  After modest editing, imaging was done
in AIPS  \citep{AIPS} using moderately robust
weighting that yielded  synthesized beams of 4.9--7.6
arcsecond in diameter.  All four sources showed CO(1--0) emission
distributions sufficiently compact to allay any concerns
about resolved-out flux.  Table \ref{TableCO} presents line flux and
velocity width measurements derived from integrated spectra
of the four galaxies.
  We also include the data for  two additional
 targets in common with the CARMA sample of  \citet{LBACO}. 
 For SDSS015028, observed with both the PdBI and CARMA, we adopt the 50$\%$ higher flux from CARMA
 (which observed the source for longer) but we adopt  the width derived from our fit, since  \citet{LBACO} 
 did not published the widths of the lines.

\begin{figure}
   \centering
  
\includegraphics[angle=0,width=8.5cm,height=8.5cm,trim=2cm 3cm 2cm
3cm,clip]{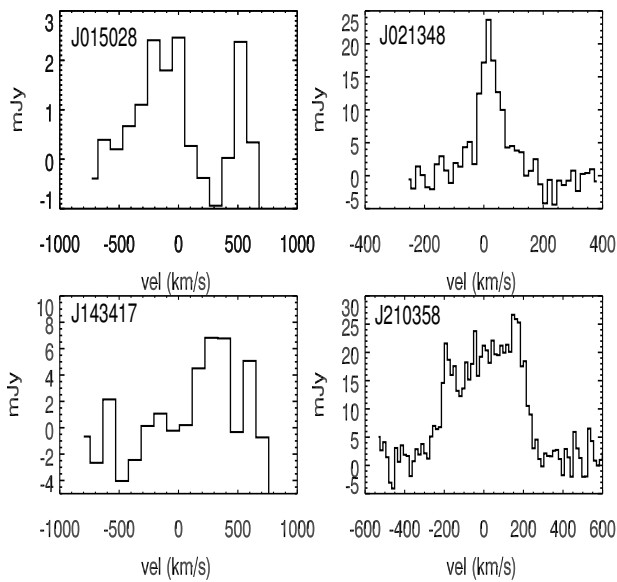}
        \caption{ IRAM PdBI CO(1-0) spectra of four galaxies.}
         \label{CO}
   \end{figure}

\noindent 
Figure \ref{CO} shows the IRAM PdBI CO spectra.
The line velocity peaks of [CII] ([OI]) and CO  agree well
for all  galaxies except for SDSSJ143417, where the CO profile  peaks at a higher velocity than
 the [CII] profile.  \citet{Basu-Zych09} describe the
 kinematics of this galaxy as disturbed, with multiple cores probably arising from a recent merger.
 It is thus possible that the bulk of the [CII] and CO emission arise from different kinematic components.

\section{Analysis}

\subsection{Comparison of the integrated velocity dispersions of the different gas phases}  
In this section we compare the velocity dispersion of the atomic, molecular, and  ionized gas.
Because we do not have spatially resolved velocity dispersion maps of LBAs, 
and hence we cannot separate  rotation from   dispersion, we will compare the $integrated$ velocity dispersions of the
different gas phases.\\
We obtain the neutral gas global velocity dispersions  by correcting the single Gaussian fitted FWHM  
(figure \ref{Spectrakms} and Table \ref{Table3})
for the instrumental resolution and converting to velocity dispersion.  
In figure \ref{sigma_comparison} we compare these dispersions  to those of H$\alpha$  \citep[from Table 1 of][]{Overzier09} and CO
when available. The  dispersions of the neutral and molecular emission lines of LBAs agree
to within the uncertainties, indicating that the neutral atomic and molecular gas  phases in the LBAs share the same
kinematics. The H$\alpha$ velocity dispersions agree with these, to within the uncertainties, 
for most of the galaxies except for SDSSJ005527 and SDSSJ080844,   
for which $\sigma_{{\rm H}\alpha}$ is almost  
twice the velocity dispersion of the neutral gas. These are also galaxies with  optical tidal emission or companions (see Table \ref{Table2}). It
is thus possible that the neutral gas is more concentrated than the ionized gas  and   does not experience the  turbulence of the gas flowing between
the interacting galaxies.\\
The LBAs integrated gas velocity dispersions are comparable to those of
  LBGs  \citep{Erba} and also to the average [CII] intrinsic velocity dispersion ($\sim$ 95  $\rm{km ~s^{-1}}$)
   of a sample of MS galaxies ($M_* \sim 10^{10}$ $M_{\odot}$) at  $0.02 <z< 0.2$2  \citep{Ibar}. 
   They are significant lower  than the [CII] intrinsic velocity dispersion of a sample of ULIRGs at $z \sim 0.3$ ($\sim$ 220
   $\rm{km ~s^{-1}}$)  published by  \citet{Magdis2014}.

\subsection{Dust temperatures}
To estimate the dust temperatures in our galaxies, we use the method developed by  \citet{Magnelli14} 
(see their Sect. 3.1).
We fit the far-infrared (FIR) photometry (from PACS spectrometer and photometer) of each galaxy with the spectral energy distribution (SED) 
templates of \citet[][DH]{Dale}.
From the  best fit to the template, we   inferred the dust temperature by 
using the pairing between dust temperature and DH templates established by  \citet{Magnelli14}.
Compared to methods relying on direct single modified blackbody (MBB) fits, this technique reduces
 the bias due to differences in the rest-frame FIR photometric data available for each fit.
Nevertheless, because the pairing established by Magnelli et al. (2014) corresponds to a fit 
of each DH template with a single MBB at $z=0$, our $T_d$ estimates are still in good agreement with other studies in the literature that rely
on such single MBB fits. We are thus  able to
compare our sample with those at high $z$ on the SFR-$M_{*}$ plane, as was done in  
\citet{Magnelli14}. We list the derived dust temperatures in column 1 of Table \ref{Table7}.
  
\subsection{Dust masses and SFR determination}
\begin{figure}
\centering
  
\includegraphics[angle=0,width=8.5cm,height=8.5cm,trim=2cm 0cm 3cm 0cm
3cm,clip]{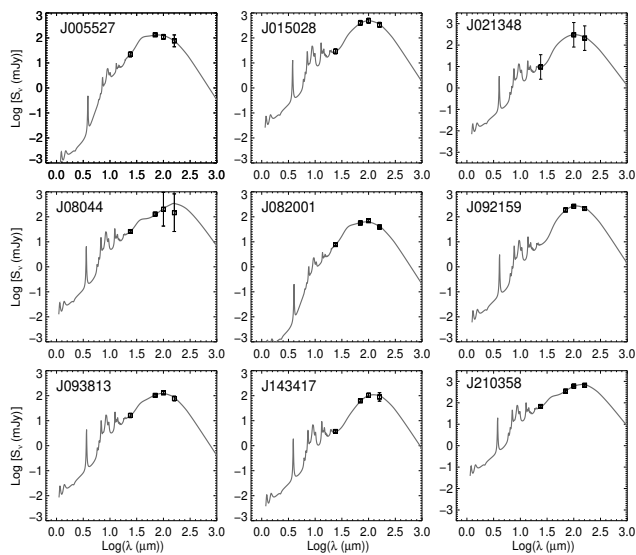}
        \caption{SED fits of LBAs FIR photometry using the method explained in Section 4.3. }
         \label{SED}
   \end{figure}

\begin{figure*}
\centering
\includegraphics[angle=0,width=14.5cm,height=8.5cm,trim=0cm 0cm 4cm 0cm,clip]{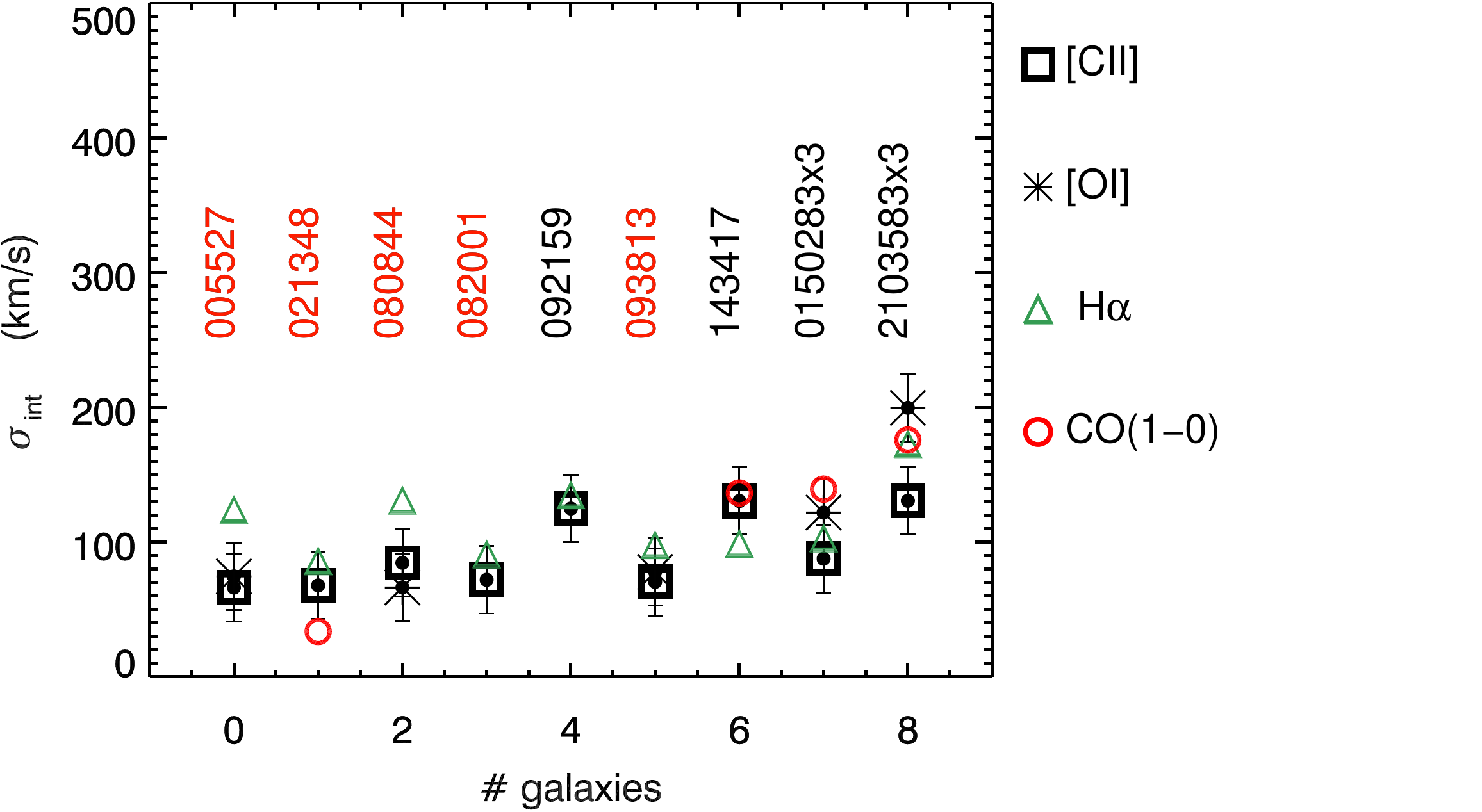}
      \caption{Comparison of the intrinsic global velocity dispersions obtained from
      the PACS [CII] and [OI] lines,  with those of CO   presented in this work  and      H$\alpha$   presented in   
      \citet{Overzier09}. Galaxies marked in red are not spectroscopically resolved. }
\label{sigma_comparison}
\end{figure*}

The IR spectral energy distributions (SEDs) of our sources were fitted
with \citet[][DL07]{Draine2007}   models,    using the {\em Spitzer} photometry at 24 and  70 $\mu$m and  the {\em
Herschel}  photometry at 
100 and 160 $\mu$m (Figure \ref{SED}). These models describe
interstellar dust as a mixture of carbonaceous and amorphous silicate
grains, whose size distributions are chosen to mimic the observed
extinction law in the Milky Way (MW), Large Magellanic Cloud (LMC), or Small
Magellanic Cloud (SMC) bar region.
In principle, the model includes six free parameters:\\
\begin{itemize}
\item[-]$q_{\rm PAH}$ parameterizes the properties of grains and essentially
defines the choice of the dust model;\\
\item[-]$U_{\rm min}$ and $U_{\rm max}$ are  the minimum and maximum stellar radiation field to which grains are exposed;\\
\item[-]$\alpha$ is the power of the dust mass dependence on the radiation field;\\
\item[-]$\gamma$ is the the fraction of dust mass locked into PDRs;\\
\item[-]$M_{\rm dust}$ is  the model mass normalization.\\
\end{itemize}

The DL07 parameter space has been limited to the range
suggested by  \citet{DraineDale} for galaxies missing sub-mm data, as in our case.
 We limit the analysis to MW dust mixtures;
$\alpha=2$; $0.7\le U_{\rm min}\le25$ and  $U_{\rm max}=10^6$. 
Finally, we use the  \citet{Li2001} values of $k_{\nu}$.
 \citet{Berta}  present a detailed discussion of  the use of DL07 models with
incomplete SEDs and a thorough discussion of the uncertainties and possible
systematics  in the determination of$M_{\rm dust}$. \\
From each fit to an  SED we also calculate $L_{IR}$(8--1000 $\mu$m), and from this the SFR following the
prescription in  \citet{KennicuttEvans}, but we divided by the factor 1.7 to convert from the Salpeter IMF
 to the Chabrier IMF that we assume in this work in order to
be consistent with  SFRs estimates for of some high-$z$ galaxy samples. 
We list in Table \ref{Table7}  $L_{IR}$ and $M_{dust}$    (with   1 $\sigma$ uncertainties)   and  SFR.
The SFRs derived in this work are a factor of ~2 smaller than those given by  
\citet{Overzier09} and listed in Table \ref{Table1}, which are calculated from H$\alpha$ and the 24 $\mu$m flux. However,  
the SFRs derived from the IR ({\it 8-1000 $\mu$}m ) luminosities  given in  \citet{Overzier11}  
are in much better agreement (within $\sim$ 20$\%$) with ours, and  the  discrepancy
can be ascribed to the different methods used to derive the IR
luminosities.  

\begin{table*} 
\caption{Line fluxes and S/N calculated from the central spaxel with point source correction for all galaxies except for those 
denoted with an asterisk, which have been extracted from the central nine (see text for details). Last two columns
list the intrinsic line and instrumental  FWHMs  at the observed wavelengths obtained from the single Gaussian fit.  
The uncertainties in the line fluxes are  $\sim$ 25$\%$  (see text for details).
}               
\begin{tabular}{clccccccccccccccccccccccccc}
\noalign{\smallskip}
\hline
\noalign{\smallskip}
 $\#$ &   Name     &         [CII]                  &    S/N      & [OI]                           & S/N     &         [CII] FWHM$_{obs}$ (instr)           &     [OI] FWHM$_{obs}$ (instr)     &    \\
      &            &    $10^{-18}$ $\rm{W~m^{-2}}$  &             & $10^{-18}$ $\rm{W~m^{-2}}$     &         &        $\rm{km ~s^{-1}}$       &      $\rm{km ~s^{-1}}$   &  \\        
\noalign{\smallskip}
\hline
\noalign{\smallskip}
 1 &  005527    &           11.6                    &   5.3       & 20.4                                  &     4.0  &            156.3 ( 196.1)           &         175.5 ( 159.2)              & \\         
\noalign{\smallskip}  
 2 &   015028*    &         62.5                    &   21.5      & 47.4                           &    7.5  &            206.3  (202.0)           &        287.2  ( 162.3)              & \\         
 \noalign{\smallskip}
 3 &   021348    &           9.0                    &   22.5      & ---                            & ---     &            159.7  (180.7)          &         ---          &    \\
 \noalign{\smallskip}
 4 &   080844    &          23.5                    &   12.0      & 14.5                           &   4.3   &            199.3  (216.8)           &         156.2 (17  1.3)     & \\
\noalign{\smallskip}
 5 &  082001    &         $<$ 5.0                   &    ---      & ---                            & ---     &            $<$169.6 (181.2)         &          ---                &  \\
\noalign{\smallskip} 
 6 &   092159    &          14.2                    &  14.5       & ---                            &  ---    &            194.2   (175.4)          &         ---                 & \\
\noalign{\smallskip}
 7 &   093813    &          28.1                    &  10.5       & 18.0                           &    4.3  &            166.0   (214.0)          &        183.0 (169.5)                &  \\
\noalign{\smallskip}     
 8  &  143417    &          20.8                    &   5.2       & ---                            &  ---    &            307.4  ( 192.3)          &          ---                &  \\
\noalign{\smallskip}      
 9 &   210358*    &         87.4                    &  40.5       & 76.0                           &14.0     &            307.7  (159.2)          &          402.6 (263.9)               &  \\
\noalign{\smallskip}
 \hline
\noalign{\smallskip}
\label{Table5}
\end{tabular} 
\end{table*}

\section{Results}
\subsection{Why study the neutral atomic gas in galaxies?}

Until recently, the ISM of high-redshift galaxies has been principally probed via 
their ionized  \citep[][ and references therein]{Glazebrook}  and molecular gas  
\citep{Tacconi13,Carilli}.
In particular, the molecular gas fraction and its depletion timescale are
two fundamental parameters  in determining  galaxy evolution  \citep{Tacconi10,Daddi10b}. 
However, the neutral atomic medium is another important component of a galaxy's ISM, and 
the study of its evolution over cosmic time is complementary to that of the ionized and molecular
gas.\\
The bulk of the atomic medium is traced by the  21 {\rm cm} HI line emission, which unfortunately 
is not yet generally accessible at redshift $\gtrsim$ 0.2.
The [CII] and [OI] lines are the main cooling lines of the cold neutral  medium 
(CNM; $n >$ 10 {\rm cm$^{-3}$} and $T < $ few $\times  10^2$ {\rm K}),  and therefore they 
can also be used to trace the  neutral atomic  phase. With the advent of   (sub)mm interferometers, this is now
possible up to very high $z$ ($\sim$ 6-7). \\
The CNM  is mostly concentrated  at the interfaces between 
 HII regions and their parent  molecular clouds ( Photo--Dissociated Regions or PDRs). 
This means that [CII] and [OI] emission are  related to star formation and can be used as SFR  tracers.
This connection in particular has attracted much interest in  recent years because 
the detection of the  [CII] line in high-redshift galaxies  could potentially be used to derive their SFRs. 
However, the two local  [CII] -- SFR calibration relationships  \citep{DeLooze,Herrera} have high intrinsic dispersion (~0.3 dex), and 
the FIR continuum is usually detectable in less time than the [CII] line.
Moreover, there is growing evidence that the [CII]-SFR relationship followed by very high-redshift galaxies
is different from the local relationship. SFGs   at $z > 5$ usually have  less [CII] emission for a given SFR 
({\it i.e.}, they have a low [CII]/IR ratio) than  local SFGs  \citep[][ and references therein]{Pentericci}.
Therefore, it is not clear whether the [CII]
emission is a better SFR tracer than the FIR continuum.\\ 
\noindent
On the other hand, the study of the content and  the dynamical state of the neutral gas 
that can be obtained with observations of lines
 gives complementary information on the more general state of the ISM in galaxies, 
 its interplay with the heating sources and its evolution over time.\\
 In recent years, many authors have shown  that normal star-forming galaxies (SFGs) from $z=0$ to $z\sim$ 3,
lie on  the Kennicutt-Schmidt relation  \citep{Genzel10,Tacconi10,Daddi10a,Tacconi13,Combes13}, 
while only mergers, i.e., local ULIRGs and  submillimeter  galaxies (SMGs) at $z=2-3$, depart from this relation.
However, the ISM conditions in local and high-redshift SFGs differ in several ways, with high-redshift galaxies showing 
not only higher pressure and column density but also higher intrinsic
velocity dispersion of the ionized medium  
(30-90 {\rm $km~s^{-1}$})  than  local SFGs ($\sim$5-10 {\rm $km~s^{-1}$})
 \citep{Law07,Natascha09,Cresci,Tacconi10,Wisnioski}. High velocity dispersion suggests  
the existence of large amounts of turbulence most likely due to star formation feedback  \citep{Newman12a}. 
Moreover,  these systems often host  very large ($\sim$ 1 kpc) clumps
that could be as massive as few $10^9M_\odot$  \citep{Elmegreen08,Genzel11,Swinbank,Freundlich} and be sites of powerful outflows 
 \citep{Newman12b}. The emission from these clumps is compatible with photoionization  by star formation and shocks from the
outflowing material. These characteristics could influence the  [CII] emission
in high SFGs in a different way than what is
observed in local SFGs.
  
\noindent  
In what follows, we analyze the [CII] (and [OI]) content  of LBAs  
 in order to establish a reference for planning similar observations of high-redshift galaxies.
So far, the observations of [CII] in high-redshift galaxies, even if steadily increasing in number, 
have mostly targeted very luminous galaxies (e.g. QSOs, SMGs ) and only in the last few years 
the first attempts to observe this line in normal SFGs at high $z$ have  started to yield the first clear detections
 \citep{Maiolino15}.
This work aims at characterizing the conditions in the CNM, and how it scales  with other 
fundamental properties of LBAs
in order to provide nearby analogs for  ISM studies of   high-redshift star-forming galaxies. \\

\begin{table}
\caption{Line flux and FWHM obtained from  single Gaussian plus continuum fit to the PdBI spectra and from  \citet{LBACO}.
}                                                                                                               
\begin{tabular}{llccc }                                                                                         
\noalign{\smallskip}                                                                                            
\hline                                                                                                          
\noalign{\smallskip}
       Name     &         CO(1-0)                   & CO FWHM                     & \\                     
                &   $ {\rm Jy ~ km ~ s^{-1}}$       &$ {\rm  ~ km ~ s^{-1}}$      & \\       
\noalign{\smallskip}
\hline
\noalign{\smallskip}                                                                                          
  015028$^*$       &        2.37                    & 328.1                         & \\                               
 \noalign{\smallskip}
  021348          &         1.83   ($\pm$ 0.2)      &  79.2                        &  \\                      
 \noalign{\smallskip}
  080844$^*$       &        1.74                    &                              &  \\                      
 \noalign{\smallskip}
  092159$^*$       &        1.04                    &                                &  \\                     
\noalign{\smallskip}                                                                                          
   143417          &        1.94  ($\pm$0.52)       & 321.3                        &  \\                     
\noalign{\smallskip}                                                                                          
   210358         &         10.68   ($\pm$0.72)     & 413.9                          &  \\                    
\noalign{\smallskip}                                                                                            
\hline                                                                                                  
\tiny{$^*$ Flux by   \citet{LBACO}} &   &    &  \\      
\noalign{\smallskip}                                                                                            
\label{TableCO}                                                                                         
\end{tabular}                                                                                                   
\end{table}

\subsection{Physical conditions in the  neutral atomic gas in LBAs}
In PDRs,  atomic gas cooling occurs mainly through  
by the [CII] and [OI] lines while the heating is due to the
photoelectric effect on  dust grains. Thus,  the
ratio between the total  emission from the FIR fine structure lines and the
total  infrared emission from grains
gives the total  photoelectric yield.  By comparing  the observed
ratio with that predicted by a model, one can derive fundamental  physical  
parameters of the neutral  ISM associated with the PDRs.
In the five galaxies observed in both [OI] and [CII], we can rule out shocks as the major emission mechanism, 
because the [OI]/[CII] 
ratio is much smaller than what is expected for shocks  \citep[$\sim$10;][]{Hollenbach1989}, and very similar to what is
expected  in normal PDRs. We therefore proceed to model the neutral gas FIR fine structure lines and the continuum as arising in PDRs.\\
We adopt the model of \citet{Kaufman}, which  considers a one-dimensional semi-infinite
slab illuminated from one side that produces optically thin  [CII], IR ($\gtrsim$ 30 $\mu$m) and
optically thick [OI], such that only one  side of the cloud   emits [OI] at 63 $\mu$m. 
The PACS beam   encompasses many
clouds, which  is equivalent to considering  many clouds randomly illuminated. Thus, we reduce the  [OI] emission
predicted by the model by a factor of 2. We also integrate  the IR flux under the 
IR SED produced as described in Section 4.2.  
\noindent 
Note that we do not attempt to correct the total [CII] emission for the contribution from ionized gas.
This is usually done by scaling from the [NII] line at 122 $\mu$m or, even better, from the lower energy [NII] 
line at 205 $\mu$m. Measurements of these lines  are not available for LBAs. 
Thus,  the observed [CII] is an upper limit to the [CII] arising in the neutral atomic medium. 

\begin{figure}
\centering
\includegraphics[angle=0,width=9.5cm,height=5.5cm,trim=0cm 0.0cm 0cm 0.5cm,clip]{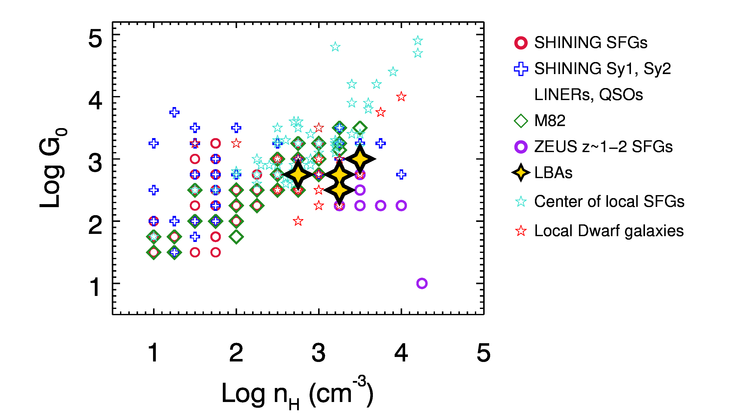}
      \caption{ LBAs on the ISRF (expressed in $G_0$ units) -- atomic gas density $n_H$ plane as inferred from the PDR modeling 
         (following  \citet{Kaufman}. 
          We also plot the results of the same modeling for the subsample of SHINING galaxies that have [OI],[CII] and FIR observations 
           \citep[][ Herrera-Camus et al. in preparation]{Javier11}, some resolved regions inside M82  \citep{Contursi},  the central regions of  local SFGs observed with ISO-LWS 
	   \citep{Malhotra,Negishi}, the sample of local dwarfs from \citet{Cormier},  and a sample of $z\sim 1-2$ SFGs observed with ZEUS 
	   \citep{Brisbin}.    }
\label{PDR}
\end{figure}

\noindent
We plot in Figure \ref{PDR} the density of the neutral medium $n_{\rm H}$ and the 
 far-UV (FUV) interstellar radiation field (ISRF) 
expressed in G$_0$ units{\footnote{  G$_0$ is the FUV (6--13.6 eV) ISRF 
normalized to  the solar neighborhood value expressed in Habing flux:
1.6$\times$10$^{-3}  \rm{erg~ s^{-1} cm^{-2}}$}}  obtained for LBAs.
We also show the results  obtained by applying  the same PDR  models  to the same set of lines and total IR emission observed in
five  other  galaxy samples:
  (1)  a local sample of  galaxies assembled   from the PACS Guaranteed Time 
Program SHINING (P.I,   E. Sturm), published in part  by  \citet{Javier11}, and in Herrera--Camus et al. in prep, which includes HII galaxies, Seyfert 1 and 2 galaxies, LINERs, LIRGs, ULIRGs and QSOs;
(2) the central regions of local star-forming, starburst, and (a few) AGN  dominated galaxies observed with ISO-LWS published by 
  \citet{Malhotra} and  \citet{Negishi};   
 (3) a sample of local low-metallicity dwarf  irregulars observed with PACS presented in  \citet{Cormier}.
 (4)  some resolved regions in the disk and the
outflow of   M82   \citep{Contursi}, and   
(5) a sample of star-forming galaxies at $z \sim 1-2$ observed with the ZEUS instrument at CSO and with $Herschel$, 
published by  \citet{Brisbin}. The latter have higher stellar  masses  and  SFRs
than our targets, but they constitute the only available  high-redshift sample of SFGs  observed in [CII], [OI], and FIR continuum 
for which it is possible to apply the same PDR modeling we have used for the  LBAs. \\
We stress that, for the  comparison samples, we do not use the published  $n_{\rm H}$ and $G_0$ values, 
but we model  the TIR continuum and  the FIR fine structure line  emission again, 
in the same way and by applying the same correction factors as  for the LBA  sample. 
This ensures a safe comparison amongst the different samples. \\

We find   10$^{2.8}  < n_{\rm H} < 10^{3.5}$ and $10^{2.5} < G_0 < 10^{3.0}$ in LBAs,
 placing them on the  high density-high $G_0$ part of the correlation,  and
 comparable to the physical  conditions in starbursts. This
is consistent with  the starburst nature of most of 
the original LBA  sample as  shown by  \citet{Overzier08,Overzier09}. 
On the other hand, the range of  $G_0$/$n_{\rm H}$ ratios obtained for the  LBAs, most of the dwarfs  and the ZEUS 
galaxies is  lower than the main relationship followed by the other galaxies. 
This could be due to  a lower $G_0$ and/or a
higher $n_{\rm H}$.  
Applying a 0.8 correction factor  to the PDR [CII] fluxes 
to account for contributions from ionized gas  \citep[e.g.,][]{Malhotra}  has negligible effect on the 
resulting $G_0$ and $n_{\rm H}$
parameters.
\citet{Brisbin} attribute this difference  to a moderately low $G_0$ 
value due to  the large, several kpc extent of star formation in their sample. 
This could also  explain  the low $G_0$ values of LBAs, since  
their UV extent is $\sim$ 5$\arcsec$  \citep[$\sim$ 5 kpc;][]{LBAHST} (see also additional 
evidence discussed in Section 6.2).
On the other hand, we cannot rule out higher gas density as a contributing factor to the 
low $G_0$/$n_{\rm H}$ values in LBAs and high-$z$ galaxies in comparison to local galaxies.\\

\begin{figure*}
\centering
\includegraphics[angle=0,width=17.0cm,height=8.0cm]{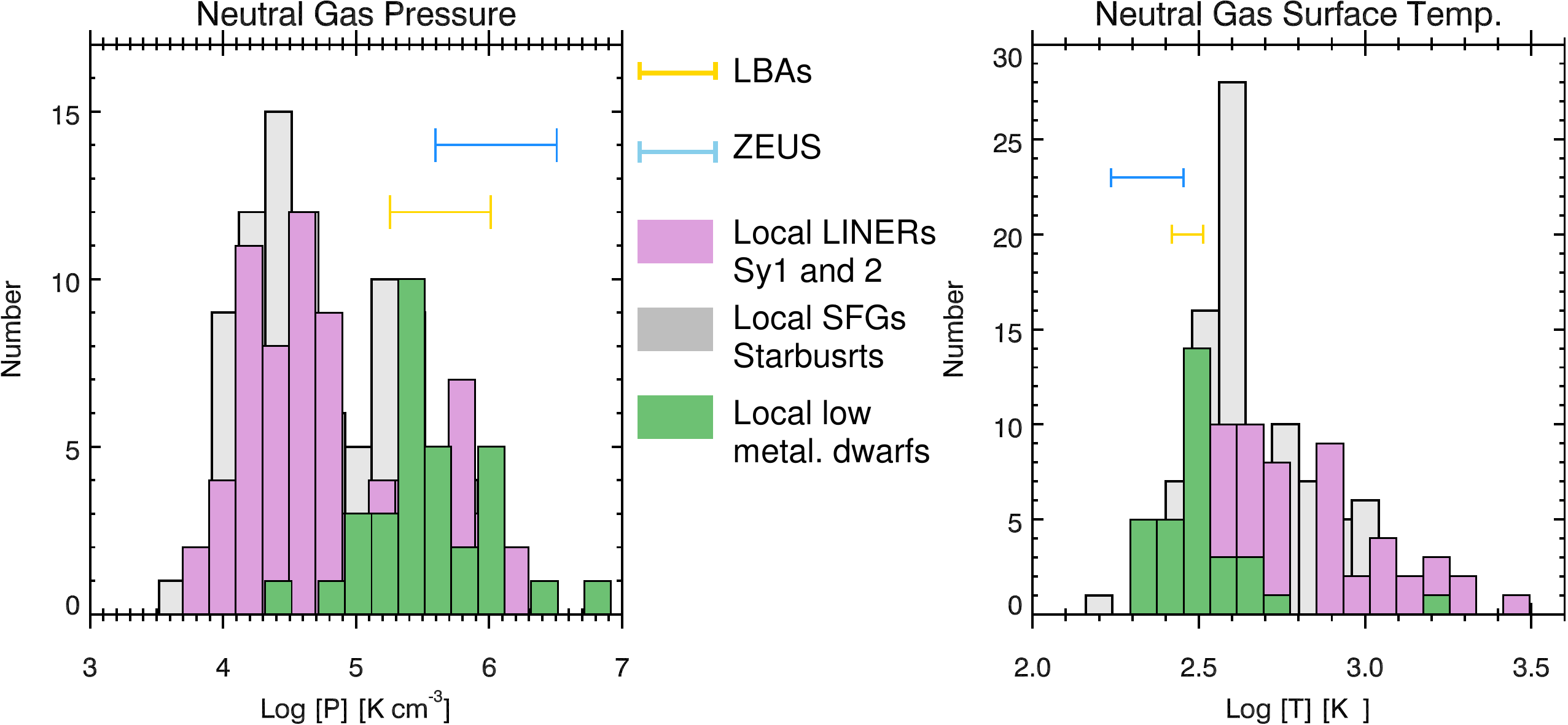}
       \caption{ Histogram of the atomic neutral gas thermal pressures and  surface temperatures 
     obtained by applying the PDR modeling to the galaxies belonging to different galaxy samples.
     The local SFGs and starbursts  belonging to the  SHINING sample, and the samples from  \citet{Malhotra} and 
      \citet{Negishi}
     are shown in gray;  the subsample of  galaxies classified as LINERs, Sy1 and Sy2 are shown in pink; and the
     low-metallicity dwarf  galaxies from  \citet{Cormier} are shown in green. 
     The ranges of ISM parameters obtained by modeling  the LBAs  and the ZEUS samples at  $z\sim 1-2$  
     are shown in yellow and   light blue respectively. }
\label{Atomic_parameters}
\end{figure*}

 \noindent 
Figure \ref{Atomic_parameters}   shows the histogram of thermal pressures and   gas surface temperatures obtained from our
modeling.
We separate the local SFGs and starbursts (shown in grey) from the local LINERs, Seyfert 1 and 2 galaxies
present in the SHINING, Malhotra et al. and Negishi et al. samples  (pink), and the low metallicity dwarf  sample (green).
For the  LBA and ZEUS samples, we   indicate  the ranges covered by these parameters, 
since the number of galaxies in each sample is small.   \\  
\noindent
LBAs and the high-$z$ galaxies of  \citet{Brisbin} have  pressures  and temperatures   
that are respectively on the high and low sides of the  histogram  ranges. 
The distribution of the pressures of local SFGs and starbursts shows a bimodality, not present in the temperature distribution,
 with  the high-pressure second  peak  dominated by starbursts.
This second peak  overlaps with the values found in dwarfs and corresponds to the range spanned by LBAs and high-$z$ SFGs. 
The thermal pressures of the cold neutral medium derived  for LBAs are   within the range 
of thermal pressures derived by  \citet{Overzier09} from the ionized gas for the whole LBA sample. 
These pressures are much higher than   the  mean thermal pressure of the CNM in our galaxy ( $\sim 3800$ {\rm K cm$^{-3}$}), 
and of atomic dominated regions in nearby galaxies  \citep{Jankins,tp}, and higher than
those derived in the nuclei of the Antennae  from X-ray  observations  \citep{Fabbiano}.
By assuming that  thermal pressure is about 10$\%$ of the total hydrostatic pressure  \citep{Cox}, we derive a total pressure of about
5$\times$$10^6$ {\rm K cm$^{-3}$} that is in the range of the total hydrostatic pressure at mid plane 
measured in high-$z$ main sequence 
galaxies  \citep[0.3-10$\times$$10^{7}$ {\rm K cm$^{-3}$}, ][]{Swinbank11,Genzel10}. \\ 
We conclude that {\it the physical conditions of the neutral atomic gas in LBAs are quite extreme with 
respect to the mean properties of the local sample of star-forming galaxies, exhibiting lower  atomic gas temperatures, 
higher pressure, and likely   higher densities,   
as found in  low-metallicity  dwarfs and starbursts. These same conditions are also seen in the
small sample of  high-$z$ star forming galaxies observed to date.}\\
\noindent
We note that hosting a DCO does not necessarily imply the most extreme ISM conditions. 
In fact, the only galaxy hosting a DCO  observed in both  [CII] and
[OI], for which   PDR modeling is possible  (SDSSJ08044), has among  the lowest values of $P$, $G_0$ and n$_{\rm H}$, and the 
highest $G_0$/n$_{\rm H}$ ratio and $T$. DCOs are very similar to the massive clumps detected in high-$z$ star 
forming galaxies  \citep{Newman12b} and therefore one would expect that their presence would be related to more extreme ISM
conditions. Unfortunately,  with our limited number of objects, it is difficult to draw 
conclusions  about a potential connection    between the conditions of the ISM and the presence of a DCO, 
by analogy with the clumps observed in high-$z$ SFG  disks.

\begin{table*}
 \caption{Dust temperatures, FIR (40--120 $\mu$m) and TIR (8--1000 $\mu$m) luminosities, and dust masses derived as described in Sections 4.1 
 and 4.2. 
 Star formation rates are derived from $L_{\rm TIR}$ using the scaling relation given by 
 \citet{KennicuttEvans},
 divided by a factor of 1.7 to convert from the Salpeter to the Chabrier IMF
 that we adopt in this work. }               
\begin{tabular}{lccccccccccccccccccccccccc}
\noalign{\smallskip}
\hline
\noalign{\smallskip}
    Name     &              $T_{dust}$    &  L$_{FIR}$        &      L$_{TIR}$   &         M$_{dust}$          &         SFR            &        \\
             &                  K         &  L$_{\odot}$      &   L$_{\odot}$    &    10$^8$$\times M_{\odot}$ &   M$_{\odot}$ yr$^{-1}$ &     \\    
\noalign{\smallskip}
\hline
\noalign{\smallskip}
   005527*    &          49.5 (2.2)       &     1.10 (0.26)e+11    &   2.17 (0.10)e+11    &    0.23 (0.09)        &         21.99         &      \\
\noalign{\smallskip}                    
   015028    &           35.9 (1.0)       &     2.75 (0.31)e+11    &   4.02 (0.13)e+11    &    1.10 (0.12)         &         40.799        &       \\
 \noalign{\smallskip}                   
   021348    &           30.2 (1.0)       &     3.98 (0.61)e+11    &   5.10 (0.10)e+11    &    1.45 (0.16)        &         51.74         &      \\
 \noalign{\smallskip}                    
   080844    &           49.4 (1.9)       &     3.71 (0.75)e+10    &   8.62 (0.73)e+10    &    4.03 (1.00)        &  8.75          &         \\
\noalign{\smallskip}                     
   082001    &           44.3 (1.0)       &     8.95 (1.00)e+10    &   1.60 (0.34)e+11    &    0.21 (0.01)        &         16.23         &        \\
\noalign{\smallskip}                    
   092159    &           35.9 (1.2)       &     3.95 (0.69)e+11    &   5.07 (0.71)e+11    &    1.80 (0.41)       &         51.44          &        \\
\noalign{\smallskip} 
   093813    &           39.6 (2.8)       &     3.40 (1.30)e+10    &   6.08 (0.25)e+10    &    0.14 (0.32)        &         6.17          &        \\
\noalign{\smallskip}                    
    143417    &           29.3 (1.0)       &    7.70 (1.40)e+10    &   1.08 (0.49)e+11    &    0.79 (0.27)        &         11.01        &          \\
\noalign{\smallskip}                    
   210358    &           33.4 (1.0)       &     2.54 (0.40)e+11    &   5.11 (0.12)e+11     &    7.2 (1.10)     &       51.90    &       \\

\noalign{\smallskip }
 \hline
\noalign{\smallskip}
\label{Table7}
\end{tabular} 
\end{table*}

\subsection{The correlation of [CII] with other ISM tracers}
\begin{figure*}
\centering
\includegraphics[angle=0,width=18.5cm,height=16.0cm, trim=0cm 0cm 0cm 0.0cm,clip]{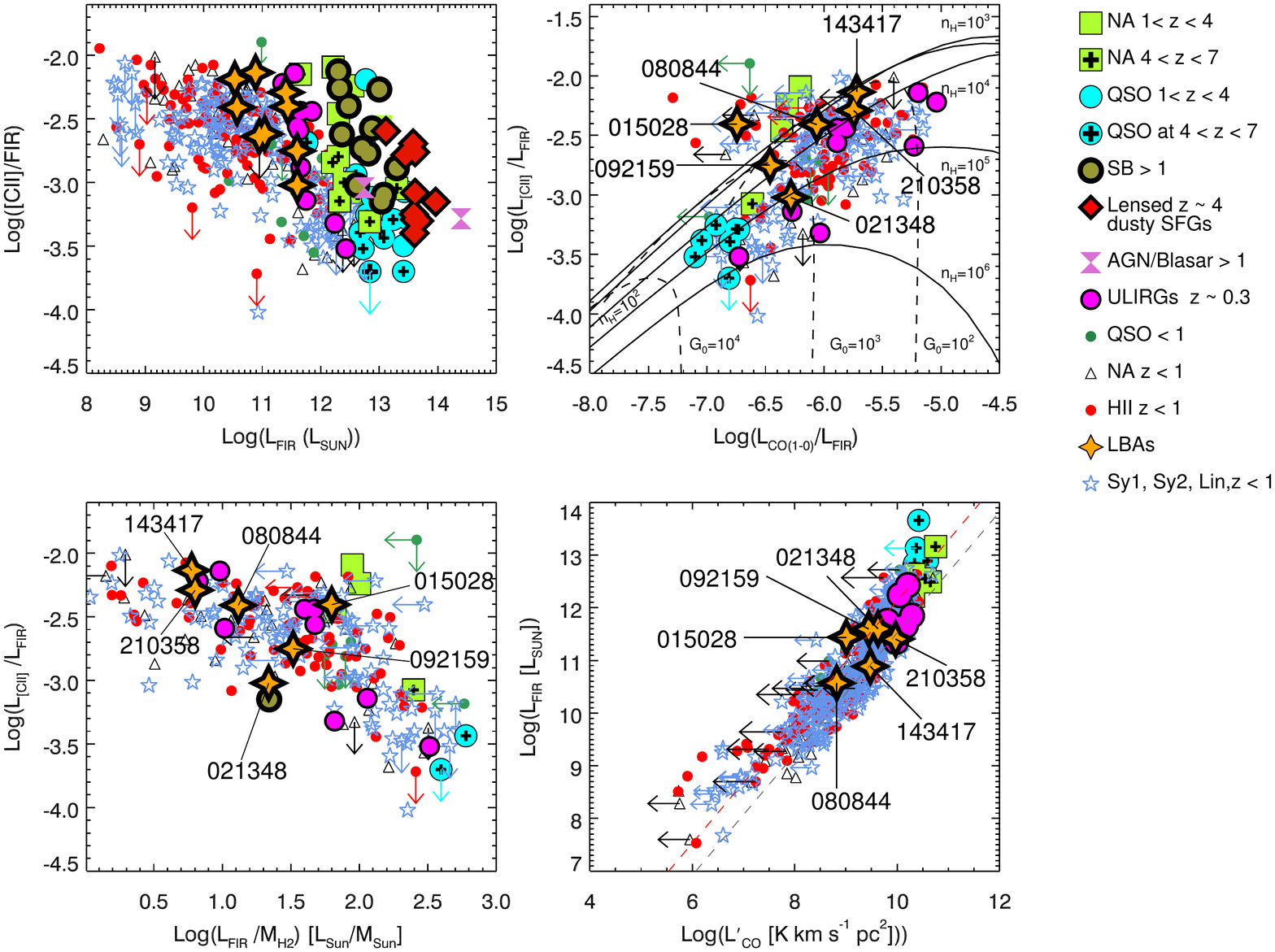}
      \caption{
      {\it Upper left panel}:  [CII]/FIR ratio  \citep[defined as in ][]{Helou}   as a function of FIR luminosity for
      the LBA sample, a set of local galaxies  belonging to the SHINING sample,  and data collected in the literature 
    \citep[Herrera-Camus et al. in preparation, ][]{Zhao}. High-redshift
      galaxies are taken from  \citet{Carilli} supplemented by data from  \citet{Wang},  \citet{DeBreuck}, \citet{Schaerer}, 
       \citet{Banados}, \citet{Yun}, and  
      \citet{Bothwell}. The sample of ULIRGs at
      $z$ $\sim 0.3$ is from  \citet{Magdis2014}.  
      {\it Upper right panel}:  [CII]/FIR ratio  as a function of  CO(1-0)/FIR
      ratio for all galaxies included in the previous panel with available CO data.
      We also plot the PDR  model grid  for constant G$_0$ and n$_{\rm H}$ 
       from  \citet{Kaufman}.
      {\it Lower left panel}: [CII]/FIR ratio as a function of $L_{\rm FIR}/M_{\rm H_2}$ ratio for the galaxies with  available 
      CO(1-0)  observations. 
      {\it Lower right panel}:  FIR luminosity as a function of   CO luminosity. 
      We also plot the relationships followed by local and high-redshift star forming galaxies (gray dashed line) and  
      by local ULIRGs and distant SMGs (red dashed line), from  \citet{Genzel10}.}
\label{CII_and_FIR}
\end{figure*}

In this Section we present and discuss correlations between the [CII] emission,
 the infrared dust emission, and the molecular gas emission of LBAs and compare them 
 to those found in other galaxy samples at low and high redshifts.\\
We first present the   diagnostic plot that  relates  the {\it [CII]/FIR ratio with the FIR luminosity}
(upper left  panel of Figure \ref{CII_and_FIR}). 
We include in this figure the values for the LBA sample,  the SHINING sample, a sample of low-redshift QSOs published by  
\citet{Zhao},
a sample of intermediate-$z$ ($z \sim0.2$) ULIRGs published by  \citet{Magdis2014}, and  a sample of
high-$z$ sources collected  by  \citet{Carilli} with recent updates (see figure caption for details).   
The high-$z$  sample   includes relatively recent high and very high-redshift [CII] detections, which have  become
more numerous thanks to new  very sensitive interferometers. However, these objects   are still biased towards
 very luminous ($L_{\rm FIR} > 10^{12}~ L_{\odot}$) galaxies (typically QSOs and SMGs) even when the targets are lensed systems.
Nevertheless, these recent observations of high-$z$ systems shed  new light on the [CII]-FIR relation, 
vastly expanding its characterization in comparison with  what was known 
only a decade ago when the diagram was limited to very nearby ($z < 0.1$) systems. Figure \ref{CII_and_FIR} 
shows that the   [CII]/FIR ratio of local systems  decreases with  FIR luminosity  \citep{Malhotra,Luhman}, 
with a knee at $L_{\rm FIR} > 10^{11}-10^{11.5}~ L_{\odot}$ beyond which the decline becomes even steeper. Extrapolating this
behavior,  
we might expect galaxies with $L_{\rm FIR} > 10^{12}~ L_{\odot}$ to have low [CII]/FIR ratios.
Observations of high-$z$ very IR-luminous  galaxies show that this is not the case: many   high-$z$ very FIR-bright sources show    [CII]/FIR ratios 
comparable to those of local, less luminous galaxies. This was first shown by 
 \citep{Maiolino09,Steve10,Stacey} 
and confirmed later by many other authors
\citep[][and reference therein]{Maiolino12,Wagg,Gallerani,Venemans}.
The high-$z$ galaxies   tend to lie in  a  sequence  parallel to that 
observed at low redshift, shifted to higher FIR luminosities. In this sequence, the objects with confirmed AGNs tend to have
lower [CII]/FIR ratios than those without AGN contributions  \citep{Steve10}. 
Thus, at high $z$ there are galaxies with luminosities as high as or higher than those of local ultraluminous galaxies, but  with   "normal" [CII]/FIR ratios. \\
LBAs follow a  correlation similar to that of the low-luminosity systems,  exhibiting a range in [CII]/FIR 
ratio decreasing with $L_{\rm FIR}$ that however never  reaches the range  of   [CII]-deficient galaxies. 
{\it  LBAs are not [CII]-deficient, and  
it is likely that  $z\sim 2$ star forming galaxies also have  normal  [CII]/FIR ratios, enhancing the likelihood that we can study their [CII] properties
with ALMA} (see Section 6.4).\\
The upper right panel of Figure \ref{CII_and_FIR} shows the [CII]/FIR ratio as a function of CO/FIR ratio and the  $n_{\rm H}$ 
and $G_0$ grids
of the Kaufman et al. model. With the exception of SDSSJ015028, which has a particularity low CO/FIR ratio, 
LBAs overlap the local SFGs but lie in ranges of $n_{\rm H}$ and $G_0$  higher than what we  obtain  by modeling 
the [CII] and [OI] lines.
This is likely due to the fact  that we are using a different pair of lines and that the Kaufman et al. model requires 
as input the total IR luminosity and not the FIR luminosity used in this diagnostic.   

\noindent

In the lower left panel of Figure  \ref{CII_and_FIR}, we show the {\it  [CII]/FIR ratio as a function of the $L_{\rm FIR}/M_{\rm H_2}$ ratio}
for galaxies with [CII] and  CO(1-0) data (the latter used to estimating   H$_2$ masses) available. 
This diagram was first shown  by \citet{Javier11}.
The values of $L_{\rm FIR}/M_{\rm H_2}$  for LBAs  have been obtained using a Milky Way conversion factor 
to derive $M_{\rm H_2}$ from the CO luminosity 
  ($\alpha_{\rm CO} = 4.3$  $M_{\odot}$ ( K  (km   s$^{-1}$  pc$^2)^{-1}$). If we   use a ULIRG  conversion factor,
  \citep[$\alpha_{\rm CO} = 0.8$  $M_{\odot}$ ( K  (km   s$^{-1}$  pc$^2)^{-1}$,  )][]{Downes}, the derived H$_2$ masses would be smaller and 
  the  $L_{\rm FIR}/M_{\rm H_2}$ ratio larger, such that
  for some galaxies it would exceed the threshold of the high-efficiency SF mode. However, as we discuss in Section
  6.1, our analysis
  favors the Milky Way conversion factor for LBAs.
  We   note that of the four LBAs with  DCOs,    
   only SDSS021348  falls below the tight correlation, although this is not the case  if a ULIRG  CO-H$_2$ conversion  
  factor is used.\\ 
The dichotomy between very luminous and less luminous systems shown before  
is less evident when the {\it [CII]/FIR ratio is plotted as a function of  $L_{\rm FIR}/M_{\rm H_2}$}.
In  galaxies in which $L_{\rm FIR}$ is predominantly powered by star formation, 
the  $L_{\rm FIR}/M_{\rm H_2}$ ratio is, to first approximation,
 the ratio between the energy released by the star formation  and the gas reservoir 
from which the stars form, and therefore is a parameter   directly related to the
star formation efficiency \citep{Eduardo,Eduardo2015,Javier11}. 
On this diagram, low and high-redshift galaxies follow the same correlation. 
This suggests that the physical mechanism responsible for the  different
[CII] emission regimes is the same locally and in the distant Universe. \\
\citet{Javier11} find that the far-infrared fine-structure line deficit 
(relative to the FIR) increases as the star formation efficiency (L$_{\rm FIR}$/M$_{\rm H_2}$) rises. They
attribute this to higher ionization parameters, resulting in the absorption of a higher fraction 
of the available UV photons by dust, leaving fewer photons to ionize and heat the gas 
 \citep{Luhman,Abel2009,Fischer,Goicoechea}.  
\citet{Javier11} and  \citet{Santos13} show that
galaxies with a [CII]/FIR  deficit lie above the MS. Thus, LBAs, that are above the local MS in
Figure \ref{LBA_and_the_others}, nevertheless have a [CII]/FIR ratio typical of MS galaxies. 
We discuss this point further in Section 6.2.
 The lower left panel of Figure \ref{CII_and_FIR} shows that the  LBAs for which we have CO data occupy the relatively high-[CII]/FIR
  region of the diagram with $L_{\rm FIR}/M_{\rm H_2}$ ratio lower than the threshold found by  
\citet{Javier11},  $\sim 80$ $L_{FIR}/M_{H_2}$. 
This is not what one might have expected in regions of high sSFR, such the clumps in DCOs.  \\

\noindent
The last relationship that we analyze is {\it between   $L'_{CO}$ and  FIR luminosity} 
(lower right panel of Figure \ref{CII_and_FIR}), in which we
 also show, plotted with red and gray dashed lines, the scaling relations for  mergers  and the star-forming galaxies that   \citet{Genzel10} 
obtain  by fitting  data for local and high redshift ($z\sim 2$)  galaxies.
Most of the  six LBAs for which we have CO data are consistent with typical MS SFEs.
The only exceptions  are  SDSSJ015028 and  SDSSJ021348 (a DCO), which lie on the merger sequence:
 for a given   amount of molecular gas, they have  FIR emission and hence SFR 
much higher than that of a star-forming galaxy. 
SDSSJ14347 and SDSSJ015028 are probably interacting systems based on a kinematic integral field spectroscopic study
of their ionized gas  by \citet{LBAHST}, but only the latter shows   
the "high"efficiency mode  of star formation typical of
merging systems. Note that the PACS beam is large enough to include both interacting components. On the contrary, SDSSJ021348, 
which contains a DCO, is on the merger
sequence even though  no clear signs of merging or interactions have been detected in the ionized gas or the UV light distribution.\\ 
We conclude that {\it most LBAs exhibit an  efficiency in turning gas into stars similar to that of MS galaxies, despite the fact
that they are starbusts and lie above the local MS.}\\
Finally, we have investigated if the [CII]/FIR and LIR/CO ratios depend  on galaxy morphology, {\it i.e.}, whether galaxies are
isolated or in some stage of interaction/merging (see Table \ref{Table1}). 
There is a tendency for the three isolated galaxies   to have
lower ratios than the others, although the small sample size prevents
us from making a strong statement.

\subsection{[CII] and [OI] as  SFR  tracers} 
\begin{figure*}[ht!]
     \begin{center}
              
\includegraphics[width=0.85\textwidth,height=0.45\textwidth]{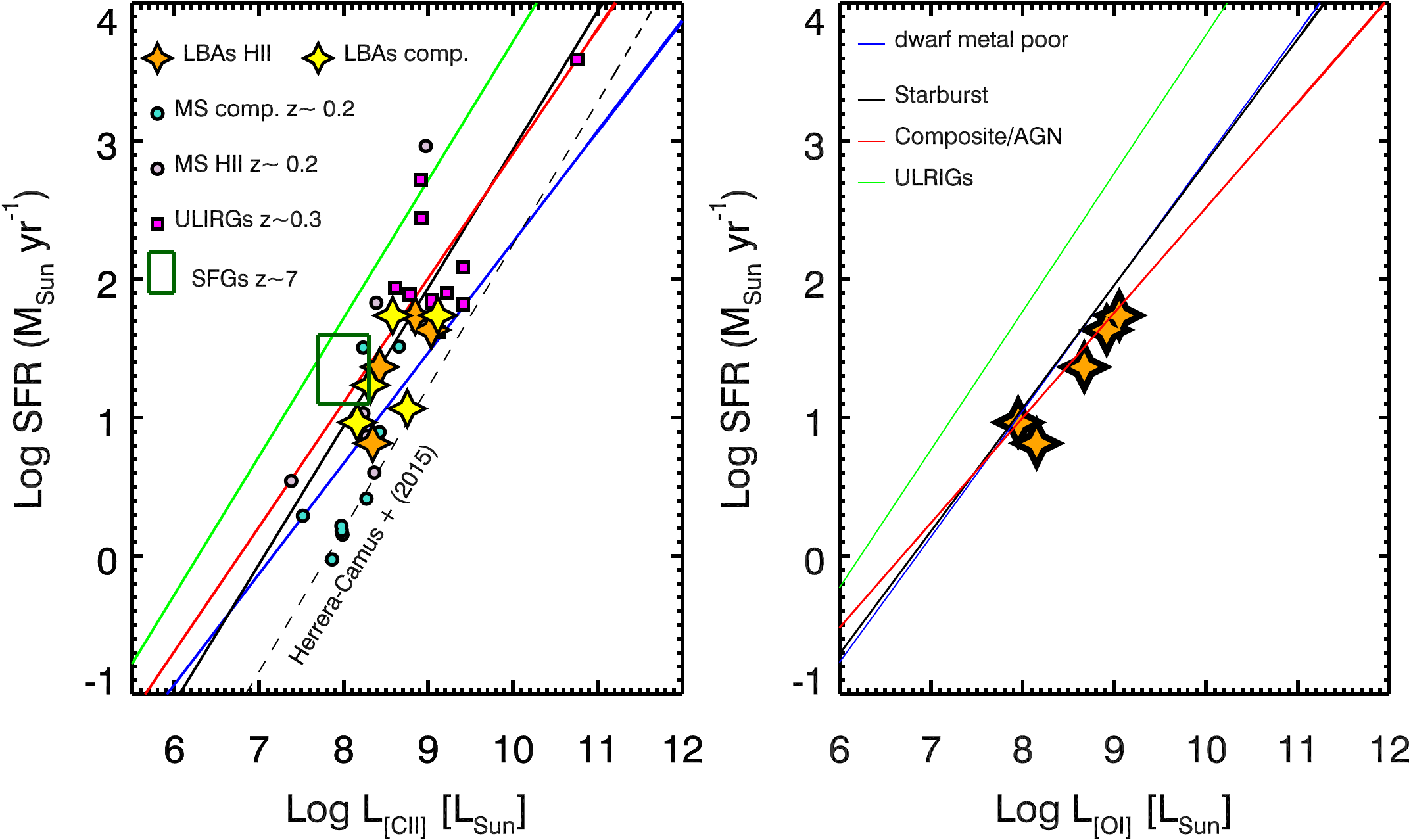}
     \end{center}
        \caption{  The relationship of  SFR with   [CII] (left panel) and [OI] (right panel) luminosity of LBAs. In the left
        panel we add samples of massive MS ($M_* \sim 10^{10}$ $M_{\odot}$) galaxies at $z\sim 0.3$  from  \citet{Ibar} 
	and of ULIRGs 
        at $z\sim 0.2$ from  \citet{Magdis2014}. In the LBA  and the Ibar et al. samples, 
        we assign different colors according to  galaxy BPT classifications.
        The green box indicates the position of four MS SFGs at redshift 7 published by  \citet{Pentericci}.
     The  lines indicate the relationships found   by
       \citet{DeLooze} for  samples of different types of local galaxies (solid colored lines, as shown in the legend)  
       and the relation followed by a sample of  normal local spirals presented  in  
      \citet{Herrera}, (dashed black line).
     The SFRs of all galaxies have been derived from $L_{\rm IR}$ and a Kroupa IMF to be consistent with the 
     derivation of  \citet{DeLooze}.       }
  \label{SFRCII} 
\end{figure*}

 In this section we explore the locations of LBAs in the  SFR -- [CII] and [OI] planes and we compare them to observed relationships 
at low and high redshifts. 
The two most recent works on the local SFR-[CII] ([OI]) relationship  are  those of  \citet{DeLooze} and \citet{Herrera}. 
 De Looze and co-authors  explore these  relations for a wide sample of local galaxies spanning several magnitudes 
 in metallicity, and for a large variety of galaxy types, including dwarfs, low-metallicity galaxies, HII/starbursts, composite and 
 AGNs, ULIRGs as well ad some high--z very IR luminous  sources.
 They use  SFRs derived from the $GALEX$ FUV plus $SPITZER$--MIPS at 24 $\mu$m for the dwarf galaxy sample and   
 SFRs derived from  L$_{\rm IR}$
 for the other galaxy populations, all assuming a Kroupa IMF \citep{Kroupa}.\\
 Figure  \ref{SFRCII} shows these relationships for the [CII] and the [OI] lines.
 On the left panel,   the locations of the LBAs are plotted together with those of  two galaxy samples 
 at comparable redshift:
 a sample of MS galaxies ($M_* \sim 10^{10}$ $M_{\odot}$) at redshift $0.02 < z < 0.2$ from  \citet{Ibar}\footnote{From 
 the original sample of  \citet{Ibar}, we have excluded the galaxies classified as Elliptical and Elliptical Irregular.}, and the
 $z \sim$ 0.3 ULIRG sample of  \citet{Magdis2014}. 
 We  plot  with different colors galaxies classified as composite/AGN and HII on the BPT diagram.
 In order to be consistent with the derivation of the De Looze et al. relationships, we have   calculated
 the SFRs of all galaxies from  $L_{\rm IR}$ using a Kroupa IMF, following the prescription from Table 3 of  \citet{KennicuttEvans}.\\
 Figure \ref{SFRCII} shows that  LBAs are well below the relationship followed by ULIRGs  (green lines), which in general have a 
   significant [CII]-deficit. But it also shows that there is a high spread in the values of all samples, 
   and that   galaxy locations   are not always consistent with their BPT classifications.
   Most  LBAs are above the relationship followed by  local  dwarfs. The
   two LBAs with  lower SFRs  closer to the dwarf relationship,  SDSSJ093813 and SDSSJ080844,
   have the lowest and the highest metallicity of the sample,  respectively (Table \ref{Table1}).  \\   
  \citet{Herrera} explore  the SFR--[CII] relation   for the KINGFISH 
 sample of local star-forming galaxies, both within each galaxy and globally. 
 They emphasize that it is better to use the 
 relation between the surface brightnesses of SFR and [CII] rather than the luminosities. 
 Since LBAs are point sources for the PACS beams we  cannot use this approach.
 Instead, we use their equation 3   relating  SFR and  [CII] luminosities. This relation 
 is shown as a dashed black line on Figure  \ref{SFRCII} after correction of 
 SFR values from the Salpeter to the Kroupa IMF.
 This relation is lower by a factor of $\sim$ 4 in SFR with respect to 
 that derived by De Looze et al. for HII/starburst galaxies.
 This is probably due to a higher percentage of starburst systems in the HII/starburst sample of  
  De Looze  et al. relative
 to the more normal  star-forming galaxies in the KINGFISH sample 
 \citep[see figure 10 in][]{Herrera}. 
 The relation falls too low to match most   LBAs and most memebrs of the other two galaxy samples.\\
The green box in figure \ref{SFRCII} represents the location of MS star-forming galaxies with stellar masses 
between $\sim 10^9-10^{10}$ $M_{\odot}$ at $z \sim 7 $ from   
 \citet{Pentericci}. They show that  for a given SFR, the [CII] emission is lower than 
what is predicted by the local SFG relationship. This is not the case for LBAs, which although with  high dispersion, 
are closer to the local starburst relationship.  
We conclude that in LBAs and in SFGs at $z \sim$ 1-2,  [CII] can be used as a rough SFR estimator. However, 
the fact that  FIR continuum  emission is detectable in less time than the FIR fine structure lines  makes 
the  use of  [CII] as a SFR  tracer  less attractive. \\

\subsection{Summary of   main results}
In this section we have presented the following results.  
(1) LBAs do not show a [CII] line deficit relative to   $L_{\rm IR}$ or $L_{\rm FIR}/M_{\rm H_2}.$ 
(2) The physical conditions in their neutral atomic gas are different from those of   
local SFGs, and more similar to those found in local starbursts, low metallicity dwarfs, and high-$z$ galaxies in terms
of thermal pressure, temperature, and $G_0$/$n_{\rm H}$ ratio. 
There is some evidence that this difference is independent of the presence of a DCO.
(3)   LBAs  do not overlap with starbursts in the density-FUV ISRF diagram, but rather 
 with local low-metallicity dwarfs
and with $z\sim$ 2 SFGs observed with ZEUS at CSO.
(4) Most  LBAs have SFEs similar to those of MS galaxies, despite their high sSFRs.
(5) LBAs roughly follow the local  HII/Starburst  SFR-[CII] relationships.  This means that the
[CII] emission in LBAs and in similar galaxies at high redshift  can be used to estimate their SFRs,
although with  uncertainties of order of a factor of 2-3.

\section{Discussion: the neutral ISM properties of LBAs in the context of galaxy evolution }

\subsection{Total molecular gas fraction}
In  recent years, many authors have shown  that the molecular gas fraction of  star forming galaxies in a given stellar mass range,
 defined as the ratio of the total H$_2$ mass 
to the sum of the stellar plus molecular gas mass,  
 increases  from 0.08 to0.47  as the redshift increases  \citep{Tacconi13}, with a possible
flattening at higher redshifts \citep{Saintonge,Magdis}, although 
the flattening is not yet confirmed  \citet{Bethermin}.
The question we  address in this section is whether our targets have a high gas fraction, strengthening also in this respect their
analogy with SFGs  at  $z \sim 2$.\\ 
Using different methods, we have calculated the total molecular gas fraction for LBAs  for which we have CO(1-0) 
data. We estimate   the H$_2$ mass with three different
CO-H$_2$ conversion  factors: one used for ULIRGs ($\alpha_{\rm CO}$ = 0.8 $M_{\odot}$ $pc^{-2}$ ($K$ 
{\rm km s$^{-1}$)$^{-1}$), one for the Milky Way ($\alpha_{\rm CO}$ = 4.3 $M_{\odot}$ $pc^{-2}$ ($K$ 
{\rm km s$^{-1}$)$^{-1}$), and one depending on  gas metallicity $Z$ as given by 
\citet{Genzel12}:  log($\alpha_{\rm CO}$) = -1.27$\times$ $Z$ + 1.18. We also calculate  the H$_2$
masses from the dust masses  listed in Table \ref{Table7}, by using a dust to gas mass ratio formula verying with metallicity 
given by \citet{Genzelscaling}.\footnote{In order to be consistent with the dust to gas mass ratio we use,  
the metallicity  should be derived from  the method given in  \citet{Pettini}, which is the case for the
metallicities we
use for the LBAs as given in  \citet{Overzier09}.} 
This formula calculates the dust to {\it molecular} gas ratio by assuming that in the  galaxies
 the molecular gas dominates  the atomic phase. The column density values for the atomic
to molecular gas transition is $\Sigma_{trans} \sim$ 14 $M_\odot$ pc$^{-2}$ \citep{Bigiel}. 
We can derive the average $\Sigma_{\rm HI}$ for three LBAs (SDSS0150283, SDSS093813, and SDSS2103583)
 for which we have PDR model results and FIR sizes from \citet{Lutz}. We calculate the mass of the atomic hydrogen
 associated with the [CII] and [OI] emission by applying :  

\begin{equation}
\frac{M_{\rm [CII]}({\rm H})}{M_{\odot}} =1.34\times \left(\frac{ L_{[\rm CII]}}{L_{\odot}}\right) \times \left(\frac{1+2 e^{(-92/T)}+n_{crit}/n_H}{2 e^{(-92/T)}}\right),  
\end{equation}
\noindent
\citep[see][for details]{Annemieke}, where we have linearily scaled the $\chi_{C^+}$ abundance by the ratio between the
solar metallicty assumed by \citet{Pettini} ($Z_{\odot}$=8.66) and the metallicity listed in Table \ref{Table1}.
We obtain  $\Sigma_{\rm HI}$ equal to  113, 15 and 159 M$_\odot$
pc$^{-2}$ respectively, close or above  the transition threshold. Hence, we can safely assume that the  dominant gas phase in LBAs
is molecular and derive gas masses from dust masses.
\\ 
We find an average gas fraction similar to that of local SFGs $<f_{gas}>= 0.08 $ only when using 
the ULIRG  conversion factor. In all the other cases,  the derived gas fraction are   $<f_{gas}>=(0.3-0.4-0.5$)
depending on the adopted  conversion, {\it i.e.} ,close to the typical gas
fraction found in high redshift galaxies \citep{Tacconi13}. \noindent
\citet{LBACO} have presented CO(1-0) data for another sample of LBAs with which we have  
four galaxies in common, and they reach similar conclusions.

\subsection{LBAs and the main sequence} 
As already mentioned in Section 2.1 and shown in Figure \ref{LBA_and_the_others}, LBAs lie above 
the main sequence  at $z = 0$. 
In this section we address the following question:  do LBAs have the typical ISM properties of local galaxies
above the main sequence, or do they behave as typical high-redshift main sequence galaxies?
In order to answer this question,  we analyze quantities tracing different 
LBA  ISM properties as a function of  distance from the main sequence at $z=0$, 1, and 2  and we compare the 
LBAs to other galaxy samples at low and high redshifts.\\

To be consistent with our derivation of SFRs and dust temperatures, we show the distances calculated 
with respect to the MS defined in  \citet{Magnelli14} and listed in column 3 of  Table \ref{Table8}. 
Table \ref{Table8} also  lists the distances from other  MS definitions, namely those derived in   \citet{Speagle}
\citet{Whitaker12}, 
and  \citet{Magdis2012}  for  redshifts 0, 1, and 2. This table  shows that the smallest distances are obtained when the
Magnelli MS is adopted (points in figures \ref{Magnelli}  and \ref{Magdis}), strengthening our conclusion.

\begin{table*}
 \caption{Distances of  LBAs from different definitions of Main Sequences  \citep{Magnelli14,Magdis2012,Speagle,Whitaker12}  at $z$=0,1 and; 2.}               
\begin{tabular}{lccccccccccccccccccccccccc}
\noalign{\smallskip}
\hline
\noalign{\smallskip}
    Name     &      $z$  & Magnelli+14  & Magdis+12 & Speagle+14 &  Whitaker+12  &\\
 \noalign{\smallskip}
\hline
\noalign{\smallskip}
   005527    &      0    &    20.03     &   49.86   &   41.33    &   33.28       &\\
             &      1    &     3.88     &    7.66   &    3.42    &    2.94       &\\
             &      2    &     1.40     &    2.15   &    1.51    &    0.62       &\\
\noalign{\smallskip}                    
   015028    &      0    &    12.14     &   23.23   &   38.70    &    23.47      &\\
             &      1    &     2.63     &    3.57   &    2.44    &     2.48      &\\
             &      2    &     1.13     &    1.00   &    0.99    &     0.63      &\\
 \noalign{\smallskip}                   
   021348    &      0    &    10.89     &   18.59   &   39.07    &    21.56      &\\
             &      1    &     2.56     &    2.86   &    2.25    &     2.42      &\\
             &      2    &     1.11     &    0.80   &    0.88    &     0.65      &\\
 \noalign{\smallskip}                    
   080844    &      0    &     6.56     &   15.76   &   14.67    &    11.27      &\\
             &      1    &     1.27     &    2.42   &    1.16    &     1.02      &\\
             &      2    &     0.48     &    0.68   &    0.51    &     0.22      &\\
\noalign{\smallskip}                     
   082001    &      0    &    12.17     &   29.24   &    27.22   &    20.91      &\\
             &      1    &     2.37     &    4.49   &     2.15   &     1.90      &\\
             &      2    &     0.89     &    1.26   &     0.94   &     0.41      &\\
\noalign{\smallskip}                     
   092159    &      0    &     6.61     &    9.26   &    27.60   &    13.22      &\\
             &      1    &     1.85     &    1.42   &     1.38   &     1.62      &\\
             &      2    &     0.76     &    0.40   &     0.52   &     0.48      &\\
\noalign{\smallskip} 
   093813    &      0    &    10.30     &   27.93   &    27.93   &    15.15      &\\
             &      1    &     2.05     &    4.29   &     4.29   &     1.22      &\\
             &      2    &     0.61     &    1.20   &     1.20   &     0.23      &\\
\noalign{\smallskip}                    
    143417    &     0    &     1.66     &    2.50   &     6.62   &     3.32      &\\
              &     1    &     0.43     &    0.38   &     0.35   &     0.39      &\\
              &     2    &     0.18     &    0.11   &     0.13   &     0.11      &\\
\noalign{\smallskip}                    
   210358    &      0    &     5.69     &    7.42   &    24.85   &    11.35      &\\
             &      1    &     1.70     &    1.14   &     1.19   &     1.43      &\\
             &      2    &     0.68     &    0.32   &     0.44   &     0.43                 \\

\noalign{\smallskip }
 \hline
\noalign{\smallskip}
\label{Table8}
\end{tabular} 
\end{table*}

\noindent
The first parameter we consider is  {\it dust temperature}.  \citet{Magnelli14} have shown 
that   dust temperature increases with  distance above the MS and as a function of  
redshift for a sample of galaxies with a wide range of stellar masses and SFRs from GOODS-N, GOODS-S and COSMOS from redshift 0
to $\sim 2$.  
Figure \ref{Magnelli} shows   the dust temperatures of LBAs, derived in the same way as for the galaxies in
 \citet[][see Section 4.2]{Magnelli14},  as a function of the LBAs distances from the  main sequences at $z=0$, 1, and
2.   The figure also shows the region
occupied by the Magnelli et al. sample  at all redshifts (blue solid box) and that occupied 
by a sample of low-metallicity lensed star forming galaxies at $z\sim 3$ from 
\citet[][red dashed box]{Saintonge}.  LBAs could represent relatively low-mass galaxies above the local main sequence   with  
high dust temperatures but still be in rough agreement with the galaxies above 
the main sequence at $z=0$ in the Magnelli sample. If the main sequences at $z=1$ and 2 are considered, LBAs have even 
higher temperatures than the Magnelli sample, but have very similar temperatures to  those found for the  lensed star-forming galaxies at $z\sim3$ from
\citet{Saintonge}.
Although we cannot  discriminate between these two scenarios, we note that the
 metallicity range of LBAs is very similar to that of the high-$z$ lensed star forming galaxies. 
 Moreover, with the exception 
 of the very high dust temperature and metallicity galaxy SDSSJ080844 (Table \ref{Table7}), 
 dust temperature of the LBAs roughly anti-correlates with   metallicity for LBAs, in agreement with the results of Saintonge et al.  \\

\noindent
Figure \ref{Magdis} shows  other   parameters related to the ISM as a function of  
distance from the main sequence, for comparison with   the similar diagrams in  \citet[][panels $a$, $b$, and $c$]{Magdis2012},
 and 
 \citet[][panel $d$]{Santos13}.
Panel $a$  shows  
 $\alpha$$_{CO}$ calculated as a function of metallicity following the relation given in the previous
section from \citet{Genzel12}; panel $b$  shows   SFE (= $L_{\rm IR}/M_{\rm H_{2}}$); panel $c$  shows the  
$L_{\rm IR}/L_{\rm CO}$ ratio;
and panel $d$  shows the [CII]/$L_{\rm FIR}$ ratio.\\
In the first three panels, we plot: (1) the values for the LBAs calculated with respect to the main sequence at $z=0$,  
1, and 2; (2) the regions occupied by the local and high-redshift MS galaxies (blue box) 
and the local ULIRGs and high-redshift SMGs (red box) 
from  \citet{Magdis2012};  
(3) the best-fit and the $\pm 1 \sigma$ values
obtained from  Magdis et al. for their full sample. \\
When we consider   the LBA  values with respect to the main
sequence at $z=0$,  only the  $L_{\rm IR}/L_{\rm CO}$ ratio follows the fit (panel $c$). 
In particular, all the LBA  values have ratios similar to those observed in ULIRGs and SMGs.
However, as shown in figure \ref{CII_and_FIR}, 
at least two of the LBAs   lie on the $L_{\rm IR}-L'_{\rm CO}$ relationship followed by the main sequence
galaxies, and this is not reflected in panel $c$  of figure \ref{Magdis}. 
When we consider  the values calculated with respect to the
main sequence at $z>1$ these two galaxies are within   one $\sigma$   of the Magdis relation, in agreement with
the findings of figure \ref{CII_and_FIR}.\\
In panels $a$ and $b$, the values of LBAs calculated with respect to the
local main sequence are outside the  regions for both MS galaxies and ULIRGs/SMGs, 
while they agree with those 
derived taking  the   main sequence galaxies at high  redshift  as a reference.  \\

Panel $d$ of figure \ref{Magdis} shows  the [CII]/FIR ratio
as a function of distance from the main sequence, analogous to   Figure 7 of  \citet{Santos13}. 
These authors analyze a sample of local LIRGs from the Great Observatories All-sky LIRG Survey (GOALS) sample 
and show that there exists a relation between the position of the galaxies with respect to the main sequence
and their [CII]/FIR ratio.
Galaxies with  sSFR greater than 3 times the sSFR of the local MS (vertical solid line in the figure) 
are starburst systems with high efficiency  star formation   and  a
progressively low [CII]/FIR ratio. Galaxies are considered to be [CII]-deficient when their [CII]/FIR ratio 
is less than $10^{-3}$ (horizontal dotted line).
Panel $d$ of Figure  \ref{Magdis} also shows the fit obtained considering only the GOALS galaxies without known AGNs 
 (solid blue diagonal line) and its
$\pm$ 1 $\sigma$ uncertainties (gray solid lines).\\
The [CII]/FIR values of the LBAs relative to the local MS are higher than those of the GOALS galaxies 
at similar distances from the main sequence; alternatively for a given [CII]/FIR ratios, LBAs seem 
to lie farther above the MS than the GOALS sample.  \\   
In this scenario LBAs seem to be extreme starburst galaxies with anomalously high [CII]/FIR ratio 
when compared to galaxies in the GOALS sample. We note that  LBAs are also LIRGs, and therefore they match  the
GOALS sample in terms of infrared emission.  In this scenario, LBAs would also have very
low  SFE ( $L_{\rm IR}/M_{\rm H_2}$, panel $b$) for their starbursting nature.  
On the other hand, when the main sequence at $z=1$ is considered, LBAs follow the fitted
relation much better. \\

\citet{Santos13} show that [CII]/FIR ratio decreases with the compactness of the starburst traced by
the FIR surface brightness  ($\Sigma_{IR}=L_{IR}/$Area). We have measured IR sizes  as half light radii at 70 $\mu$m, for only four
 of our targets   \citep{Lutz}. \citet{Santos13} calculate   sizes using the half light radius in the
MIR ($R_{e,MIR}$). However,  the bulk of the heating 
sources responsible for the emission at 70 $\mu$m and in the MIR should be sufficiently  similar to allow  a fair  comparison. 
We find that the four LBAs for which we have reliable FIR sizes lie on the relation 
presented by these authors (their Figure 5)  with low FIR surface brightness ($\Sigma_{\rm  IR}$ <
$2\times10^{10}$ $L_{\odot}$ {\rm kpc}$^{-2}$),  indicating that LBAs likely have extended star formation, as already suggested by the analysis of the $G_0/n_H$ values in Section 5.2, 
and in agreement with their high [CII]/FIR ratio.\\
\noindent
{\it We conclude that  LBAs represent a very peculiar local population with ISM properties 
very different from those of  local galaxies in the same region of the $M_*$-SFR plane and  with the
same infrared emission.}  
 Indeed, the ISM properties of  LBAs  seem to be more similar to those of $z\sim 1-2$ 
main sequence galaxies. This  result is consistent  with the finding of Section 5.2, where we have shown
that in LBAs, the physical parameters of the CNM are similar to those found in
 high-redshift galaxies rather than in local SFGs. 
 
   \begin{figure}
   \centering
\includegraphics[angle=0,width=8cm,height=8cm, trim=4cm 4cm 11.0cm 6.0cm,clip]{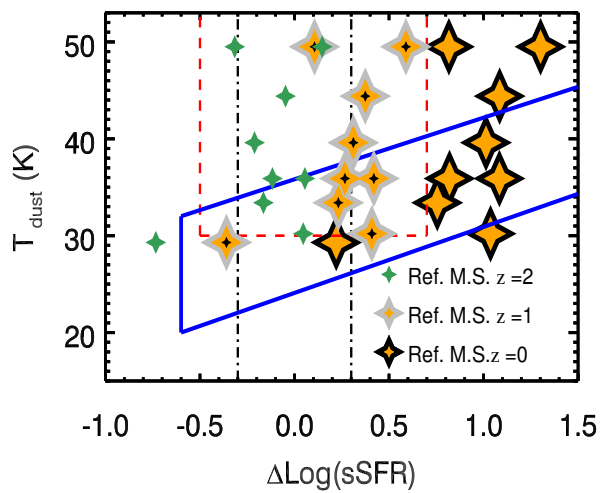}
      \caption{LBAs on the relationship between their dust temperature and their distance from 
      the main sequence at redshifts 0, 1, and 2, as indicated in the legend. We add the blue the box enclosing the
      relation  followed by a sample of galaxies from local to $z\sim2$ presented by   
      \citet{Magnelli14} and in red the region occupied by a sample of $z\sim3$ lensed low metallicity star forming galaxies  presented by  
      \citet{Saintonge}.}
        
         \label{Magnelli}
   \end{figure}

   \begin{figure*}
   \centering
  
\includegraphics[angle=0,width=16cm,height=12cm, trim=1cm 1cm 1.8cm 3cm,clip ]{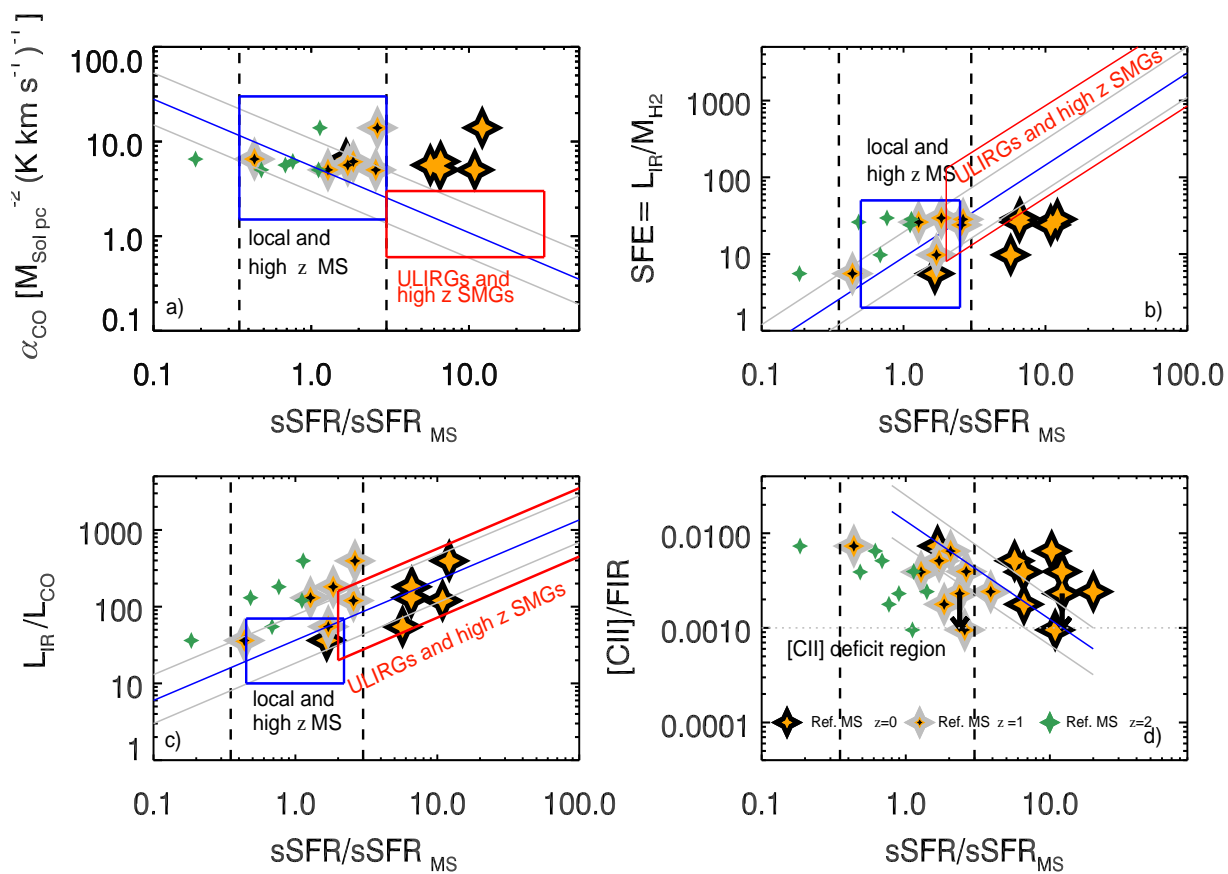}
      \caption{LBA ISM parameters as a function of  distance from the MS at $z=$0, 1 and 2 
      (using the same symbols as in Fig. \ref{Magnelli}). The
      blue solid box represents the region occupied
      by  SFGs on the MS and the red solid box that by galaxies above the MS. }
          \label{Magdis}
   \end{figure*}
 
\subsection{[CII] ALMA observations of main sequence galaxies at $z  \sim   2$ } 
We have shown that the ISM properties of LBAs are very similar to those of  main sequence galaxies at
higher redshifts. In light of this analogy  we can use the observed [CII] fluxes  of LBAs to 
predict how much time is necessary to observe [CII] in main sequence galaxies at $z$ = 2 with ALMA.\\
The expected  [CII] peak flux of an LBA   redshifted to $z=2$  ranges
from $\sim$ 10 to 50 mJy. 
This refers to the emission of the entire galaxy because LBAs are unresolved by PACS.
We assume an intrinsic galaxy extent of  $\sim$ 6$\arcsec$,  the typical
H$\alpha$ size of $z\sim2$ main sequence galaxies.
The time required  just to  $detect$ [CII]  with ALMA with  S/N $>$ 10 
and a velocity bin equal to  50 {\rm km s$^{-1}$}  is  $\lesssim$ 2 $h$ with 43 12--m  antennae.\\
Assuming the flux surface brightness decreases exponentially and matching the 70 km s$^{-1}$ 
typical velocity resolution of H$\alpha$ observations, 
the time required to $map$  a $z=2$ main sequence galaxy   is 
 12 -18 h depending whether 50 or  the current 43 offered antennae are considered.\\ 
 We conclude that  in its final state of development, ALMA will be able
 to detect a large number of relatively low-mass main sequence galaxies at $z=2$ and to map a small subsample.

\section{Summary and conclusions}  
We have presented an analysis of the neutral interstellar medium of a sample of
Lyman Break Analogs (LBAs) at $z\sim$0.2 observed in [CII], [OI],  and   FIR continuum  
with the $Herschel$/PACS instrument and in the $^{12}$CO(1-0) line with the IRAM PdBI. 
By definition, LBAs share many fundamental properties with Lyman Break Galaxies at high redshift.
All our targets are in the LIRG  regime, but they have UV/FIR ratios much higher than
LIRGs.
The main goal of this paper was to investigate whether   the ISM physical conditions in LBAs
resemble those of their high-redshift analogs and thus to offer a  unique reference sample  for
planning future submillimeter observations of high-redshifts star-forming  galaxies.
The main conclusions of this analysis are the following:\\

\begin{itemize}
\item[-]{We  have detected all galaxies in line and FIR continuum  emission. The
    atomic, molecular, and  ionized gas share the same global kinematics  as
  traced by the  integrated line  velocity dispersions.
  We have not detected outflows in the FIR fine structure lines, even in targets where  
  high-velocity ionized gas outflows have been observed  \citep{Overzier09,Heckman11}, 
  possibly because the  S/N  of the FIR lines is too low to detect relatively faint line wings.}
\item[-]{By modeling the FIR [CII] and [OI] lines and TIR dust continuum as arising in PDRs, we show that
the physical conditions of the neutral atomic gas in LBAs are quite extreme with
respect to the mean properties of local spiral galaxies: LBAs have lower
gas temperatures and higher densities and pressures, similar to the conditions
in  high-redshifts star-forming galaxies. The $G_0/n_{\rm H}$ ratios of LBAs are lower than in local starbursts, analogous to
what  has been found in some high-redshift galaxies by  \citet{Brisbin}, and 
consistent with the interpretation that  star formation in these galaxies is spread over
several kpc  and/or the gas density is higher than that in local SFGs. This  interpretation is also supported 
by the low $\Sigma_{\rm IR}$ values (that trace  starburst compactness) in at least four targets of our sample.}

\item[-]{LBAs do not suffer  from the {\it [CII]-deficit} shown by   infrared-bright galaxies,
especially those with a significant AGN contribution. This is important for
planning future observations of high-redhifts main sequence galaxies with   available submillimeter
interferometers in the [CII] line.}

\item[-]{The [CII] emission of LBAs roughly follow  the local HII-starburst [CII]-SFR relationship  
published by  \citet{DeLooze}. Despite the fact that LBAs show dense and warm gas
similar to the gas in local dwarf galaxies, with which they also share stellar masses and metallicities, 
their  [CII]/SFR ratio   is higher than in local dwarfs, suggesting that LBAs belong to a distinct local population.  
We stress however, that the SFR-[CII] relationship has  high dispersion, so SFRs cannot be derived 
with a precision better than a factor of $\sim 3$, and that the FIR continuum is a  more  efficient way  to
derive SFRs.}

\item[-]{The CO/FIR ratios of LBAs are  typical of  MS SFGs, except for two targets that lie on the  merger sequence. }

\item[-]{We derive the total molecular gas fraction in LBAs with four methods, obtaining significantly higher fractions than  in local
SFGs with all methods except when   ULIRG-like CO to H$_2$ conversion factors are used. This again strengthens
the analogy between  LBAs and high-redshift SFGs.}

\item[-]{LBAs lie above the local main sequence, but we  find that,  when
their  ISM properties are plotted as a function of their   
distance from the local MS,  they do not lie in any of the regions   occupied by other local samples.  
On the other hand, when the MS at $z>1$ is considered, LBA values fall  on the same regions occupied by high-redshift  main sequence galaxies.  
This suggests that   LBAs have ISM properties unique and never encountered in local galaxies above and on the MS, but 
  similar to those of main sequence galaxies at high-redshift. } 
 
\end{itemize}
\noindent
In summary, LBAs are distinct in their ISM properties from local spirals, dwarfs, and ULIRGs although they share
some  properties with each of these populations. Nevertheless, many ISM properties of LBAs are
similar to those found in high-redshift main sequence galaxies. Therefore, we confirm that LBAs are, as in other respects,  true
analogs of these galaxies  in their ISM properties as well.\\
\noindent
In light of this conclusion, we have calculated the observing time needed  to detect a typical Lyman Break Galaxy 
({\it i.e.}, a high redshift main sequence galaxy wit $M_* \sim 10^{9.5-11}$ $M_{\odot}$) at redshift
2  in [CII] with band 9 of ALMA. 
We find that one can detect such a target in a few
hours with S/N =10 and  map it in 12-18 hr (depending on the number of antennae).

\begin{acknowledgements}

The authros wish  to thank the anonymous referee for the very useful comments and suggestions that improved the paper.\\
PACS has been developed by a consortium of institutes led
by MPE (Germany) and including UVIE (Austria); KU
Leuven, CSL, IMEC (Belgium); CEA, LAM (France); MPIA
(Germany); INAF-IFSI/OAA/OAP/OAT, LENS, SISSA (Italy);
IAC (Spain). This development has been supported by the
funding agencies BMVIT (Austria), ESA-PRODEX (Belgium),
CEA/CNES (France), DLR (Germany), ASI/INAF (Italy), and
CICYT/MCYT(Spain). \\
A.J.B. acknowledges support from NFS grant AST-0955810.\\
A.Verma acknowledges support from the Leverhulme Trust in the form of 
a Research Fellowship $\&$ a grant from the University of Oxford's Returning Carers Fund.\\

\end{acknowledgements}

\end{document}